\documentclass[twocolumn,a4paper,amsmath,amssymb,showpacs,prb,superscriptaddress]{revtex4-2}

\usepackage{graphicx} 
\usepackage[dvipsnames]{xcolor} 
\usepackage{bm}
\usepackage[normalem]{ulem}
\usepackage{siunitx} 
\DeclareSIUnit{\muB}{\mu_{\mathrm{B}}} 
\DeclareSIUnit\angstrom{\mathring{\mathrm{A}}} 

\begin{document}

\title{Microscopic Origin of Piezomagnetism in Mn$_3$Sn: A Dual Real- and $k$-Space Picture}
\author{Soichiro Kikuchi}
\affiliation{Department of Materials Science, Graduate School of Engineering, Osaka Metropolitan University, Sakai, Osaka 599-8531, Japan}
\author{Yuki Yanagi}
\affiliation{Liberal Arts and Sciences, Toyama Prefectural University, Toyama 939-0398, Japan}
\author{Thi Ngoc Huyen Vu}
\affiliation{Institute for Materials Research, Tohoku University, Sendai, Miyagi 980-8577, Japan}
\author{Michi-To Suzuki}
\affiliation{Department of Materials Science, Graduate School of Engineering, Osaka Metropolitan University, Sakai, Osaka 599-8531, Japan}
\affiliation{Center for Spintronics Research Network, Graduate School of
Engineering Science, Osaka University, Toyonaka, Osaka 560-8531, Japan}
\date{\today}
\begin{abstract}
We present a comprehensive first-principles study on the origin of the piezomagnetic effect in the non-collinear antiferromagnet Mn$_3$Sn, a material known for exhibiting a large anomalous Hall effect. We investigate strain-induced variations of electronic and magnetic states and elucidate the mechanism of the piezomagnetic effect from both real-space and momentum-space perspectives. In real space, the emergence of piezomagnetism is understood to arise from rotations of the magnetic moments at specific Mn sites, which directly couple to the strain.
Through detailed electronic structure analysis, we identify the Fermi surfaces that play a crucial role in the emergence of piezomagnetism. Our results reveal that specific Fermi surface features undergo pseudo-degeneracy lifting under applied strain, which significantly contributes to the induced net magnetization. 
By combining these complementary real-space and momentum-space pictures, our dual-space analysis provides deep insight into the microscopic origins of strain-driven magnetization in Mn$_3$Sn.
\end{abstract}

\maketitle

\section{Introduction}

In recent years, the antiferromagnet Mn$_3$Sn has attracted significant attention across various fields, including both applied and fundamental research. Mn$_3$Sn crystallizes in a hexagonal structure and exhibits magnetic order with a non-collinear magnetic structure below the Néel temperature $T_N\approx420\rm{\,K}$~\cite{Tomiyoshi1982, Brown1990}. In this non-collinear magnetic structure of Mn$_3$Sn, the spin moments of the six magnetic Mn atoms within the unit cell largely cancel each other out, resulting in an extremely small net magnetization on the order of $\sim0.003\rm{\,\mu_B/Mn}$~\cite{Brown1990}.  Therefore, Mn$_3$Sn can be regarded as a broadly defined antiferromagnet with almost no net magnetization~\cite{Tomiyoshi1982, Brown1990, Nakatsuji2015}. Nevertheless, Mn$_3$Sn is known to exhibit many unique phenomena, including those typical of ferromagnetic materials, such as the anomalous Hall effect and the magneto-optical Kerr effect~\cite{Nakatsuji2015, Suzuki2017, Higo2018, Ikhlas2022}. 

Among the various phenomena exhibited by Mn$_3$Sn, one of particular interest is the piezomagnetic effect: the induction of a net magnetization by an applied strain~\cite{Dzialoshinskii1958,Moriya1959,BorovikRomanov1994}.
Historically, piezomagnetic effects have been first observed in collinear antiferromagnetic insulators  CoF$_2$ and MnF$_2$~\cite{BorovikRomanov1960} with  N\'{e}el temperatures $T_{\mathrm{N}}=\qty{37.7}{\kelvin}$~\cite{Stout_1953} and $\qty{66.5}{\kelvin}$~\cite{Stout_1942}, respectively. 
Subsequent studies on piezomagnetism and its inverse effect have primarily focused on antiferromagnetic insulators~\cite{Andratskii_1966, Phillips1967,Kadomtseva_JETPLett1981,Kadomtseva_JETP1981,Zvezdin_JETP1985,Aoyama_PhysRevMaterials.8.L041402, Jaime2017,nanjo2025piezomagneticeffect5dtransition} with an exception of a ferromagnetic metal URhGe~\cite{Tomikawa_PhysRevB.110.L100408}. 

Piezomagnetism has attracted further interests from various perspectives such as, a hallmark feature of altermagnets~\cite{Ma_NatCommun2021,Aoyama_PhysRevMaterials.8.L041402,Bhowal_PhysRevX.14.011019,Naka2025,Ogawa_2025} and cross-correlation in noncollinear magnets~\cite{Gomonaj_phasetransition1989,Lukashev2008,Zemen2017,Jaime2017, Ikhlas2022,Boldrin2018,Samathrakis2020,Arima_JPSJ2013,nanjo2025piezomagneticeffect5dtransition}.
More recently, its presence has been reported in Mn-based compounds such as Mn$_3$$X$ ($X$ =  Sn, Ge) and Mn$_3$$A$N ($A$ = Ga, Zn, Ni)~\cite{Ikhlas2022, Gomonaj_phasetransition1989, Lukashev2008, Zemen2017, Boldrin2018, Samathrakis2020}. 
In contrast to the typical piezomagnetic compounds, these are metallic antiferromagnets exhibiting the piezomagnetic effect at room temperatures.




The piezomagnetic effect of Mn$_3$Sn is particularly significant because it enables not only the control of magnetization but also the tuning of the sign of the anomalous Hall effect via strain~\cite{Ikhlas2022}, and various engineering applications are also anticipated~\cite{Choi2025}. 
The piezomagnetic effect in antiferromagnets typically arises when anisotropic strain is applied to a magnetic material with a specific magnetic structure, thereby modifying its magnetic symmetry and inducing a net magnetization proportional to the applied stress~\cite{Dzialoshinskii1958,Moriya1959,BorovikRomanov1994}. 
However, in certain non-collinear antiferromagnets such as Mn$_3$Sn, the piezomagnetic effect can occur without lowering the overall magnetic symmetry of the non-collinear magnetic order that allows the emergence of the anomalous Hall effect~\cite{Huyen2025}. 

Both experimental and theoretical studies have demonstrated that the piezomagnetic effect in Mn$_3$Sn originates from the rotation of the magnetic moments of certain Mn atoms under uniaxial pressure, which breaks the balance of mutually canceling magnetic moments~\cite{Meng2025, Huyen2025}. While this provides valuable insight into the macroscopic manifestation of the piezomagnetic effect, the microscopic mechanism underlying the induction of magnetization remains unclear. Building on these recent advances, we aim to elucidate the microscopic mechanism of magnetization induction in a particular antiferromagnetic order of Mn$_3$Sn by means of first-principles calculations. The reminder of this paper is organized as follows. Section II describes the computational methods, Section III presents the results and discussion, and Section IV concludes the paper.

\section{METHOD}

First-principles calculations were performed using the Vienna Ab initio Simulation Package (VASP) based on density functional theory (DFT)~\cite{Kresse1993, Kresse1994, Kresse1996, Kresse1996CMS}. The exchange-correlation energy was treated within the generalized gradient approximation (GGA) using the Perdew–Burke–Ernzerhof (PBE) functional~\cite{Perdew1996}. Spin–orbit coupling (SOC) was included self-consistently in all calculations to accurately capture the relativistic effects essential for describing the piezomagnetism in the non-collinear antiferromagnetic states of Mn$_3$Sn~\cite{Huyen2025}. A plane-wave energy cutoff of 600 eV was used, and the Brillouin zone was sampled using a $\Gamma$-centered 16$\times$16$\times$16 Monkhorst-Pack $k$-point mesh.

We obtain the optimized lattice constants $a=b=\qty{5.567}{\angstrom}$ and $c=\qty{4.432}{\angstrom}$ at zero external pressure, consistent with our previous study~\cite{Huyen2025}, and, to evaluate the piezomagnetic response under anisotropic strain, we considered a uniaxially compressed structure along the $x$ axis corresponding to approximately 0.4\% strain, which reults in lattice parameters of $a=\qty{5.545}{\angstrom}$, $b=\qty{5.566}{\angstrom}$, and $c=\qty{4.435}{\angstrom}$. For this strained configuration, atomic positions were also fully relaxed to ensure structural stability. The electronic structure and Fermi surface evolution under the introduced strain were further analyzed to identify the key factors contributing to the emergent magnetization.

\section{RESULTS}

\subsection{Real space picture}
\label{Sec:spin moment}

\begin{figure}
    \centering
    \includegraphics[width=1.0\linewidth]{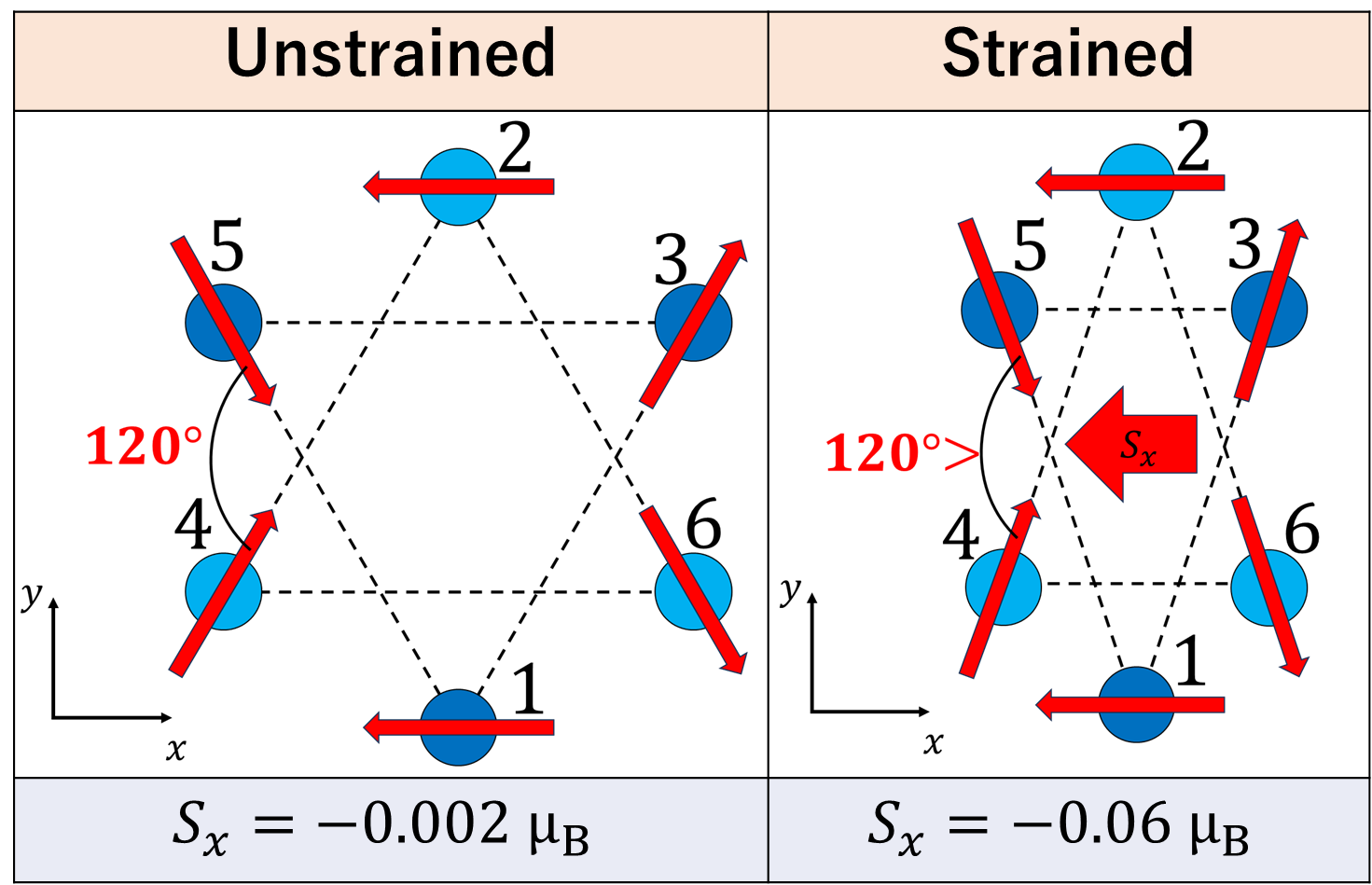}
    \caption{Change in magnetic moments under the uniaxial pressure along $x$-axis.}
    \label{fig:un-st_AFM1}
\end{figure}

To establish a foundation for the momentum-space analysis presented later, we first analyze the spin moments in real space. While such behavior has already been recognized in earlier studies, it is crucial to revisit it here in order to clarify how uniaxial compression modifies the balance of magnetic moments.
Figure~\ref{fig:un-st_AFM1} depicts the spin moments in the unstressed state and after compression.
In the unstressed state, the crystal exhibits hexagonal symmetry in the paramagnetic phase; however, the magnetic order breaks this symmetry, lowering it to orthorhombic and resulting in the magnetic point group $mm'm'$.
The magnetic point group $mm'm'$ forbids the $y$-component of magnetization, which is accordingly cancelled among the Mn3–6 atoms, as seen in Fig.~\ref{fig:un-st_AFM1}. Although the $x$-component $M_{x}$ is largely cancelled between Mn1–2 and Mn3–6, the symmetry allows a finite $M_{x}$, resulting in incomplete cancellation and a very small net magnetization.
Since the magnetic structure is already orthorhombic, the uniaxial pressure along $x$ axis does not cause further magnetic symmetry breaking, and the magnetic space group thus remains $mm'm'$. 
 Meanwhile, the uniaxial pressure affects the magnitude and direction of the magnetic moment on the Mn atoms, disrupting the balance of cancellation and inducing magnetization without breaking the magnetic symmetry.

In the unstressed state, each Mn atom carries a spin moment of about $\qty{2.90}{\muB}$ lying in the basal plane ($\theta =90^{\circ}$). Mn1 and Mn2 are aligned at $\phi = 180^{\circ}$, while Mn3–Mn6 are oriented at $\phi = \pm 60.02^{\circ}$, resulting in an almost complete cancellation and a very small net magnetization of $M_{x} = \qty{0.002}{\muB}$. When uniaxial compression is applied along the $x$ axis, the magnitude of the local spin moments decreases only slightly (to $\qty{2.89}{\muB}$), but the azimuthal angles of Mn3–Mn6 rotate by about $0.4^{\circ}$ within the $xy$ plane (to $\pm 60.39^{\circ}$). This small reorientation breaks the delicate balance of cancellation and enhances the total spin moment to $S_{x} = \qty{-0.06}{\muB}$, corresponding to a net magnetization of $M_{x} = \qty{0.06}{\muB}$. Such strain-induced rotation of Mn spins, which generates magnetization without changing the underlying magnetic symmetry, has also been reported by Huyen et al.~\cite{Huyen2025}. 

\subsection{$k$-space picture}
\label{Sec:spin dis k-spa}

\begin{figure}
    \centering
    \includegraphics[width=1.0\linewidth]{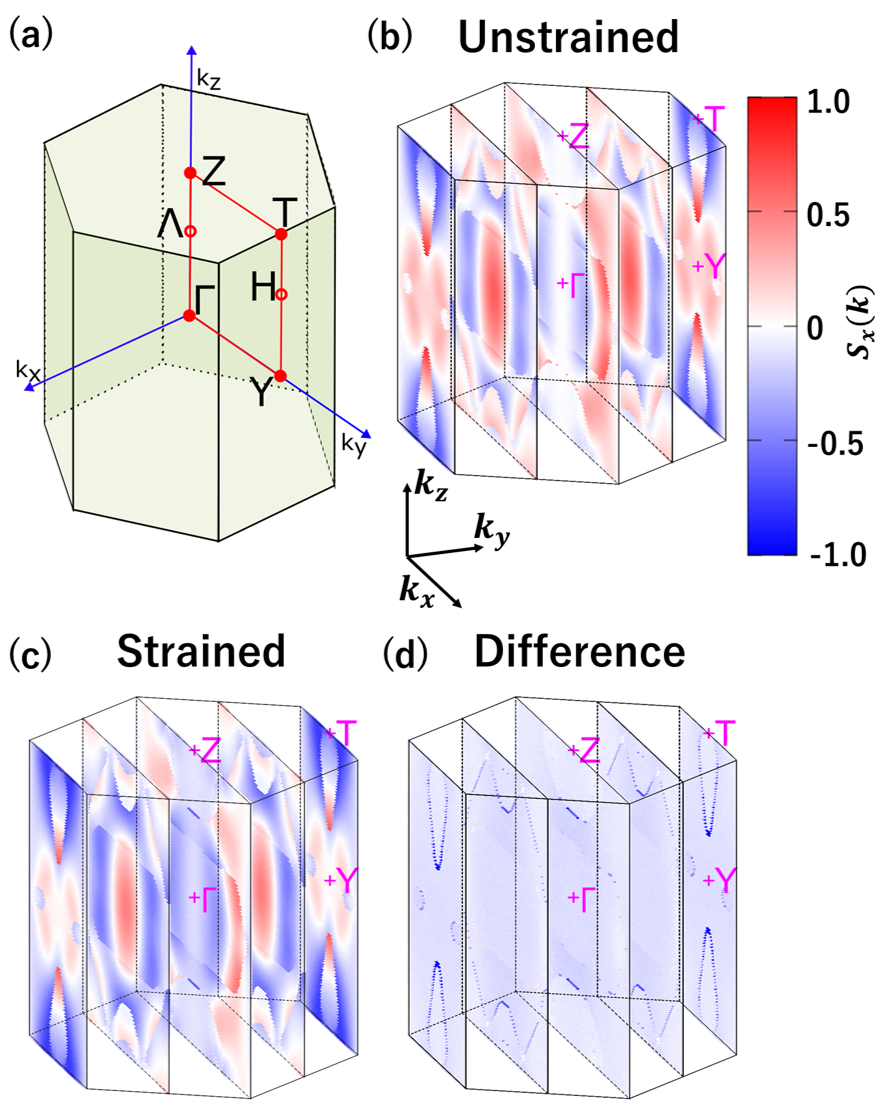}
    \caption{(a) Orthorombic magnetic Brillouin zone of Mn$_3$Sn. 
    (b)-(d) Distribution of the x component of spin, $S_{x}({\bm k})$, (b) before and (c) after compression, and (d) the difference between the two within the Brillouin zone of Mn$_3$Sn. In Panel (a) The labeling of the high-symmetry points and lines in the Brillouin zone follows the convention used for base-centered orthorhombic lattices~\cite{Bradley_book_2010}.Panel (d) is obtained by mapping the compressed Brillouin zone onto the uncompressed one through reciprocal-lattice scaling, expressing $S_{x}(\mathbf{k})$ in the common $\mathbf{k}$ coordinates, and plotting their difference.
     }
    \label{fig:BRZ+ST3D}
\end{figure}

\begin{figure}
    \centering
    \includegraphics[width=1.0\linewidth]{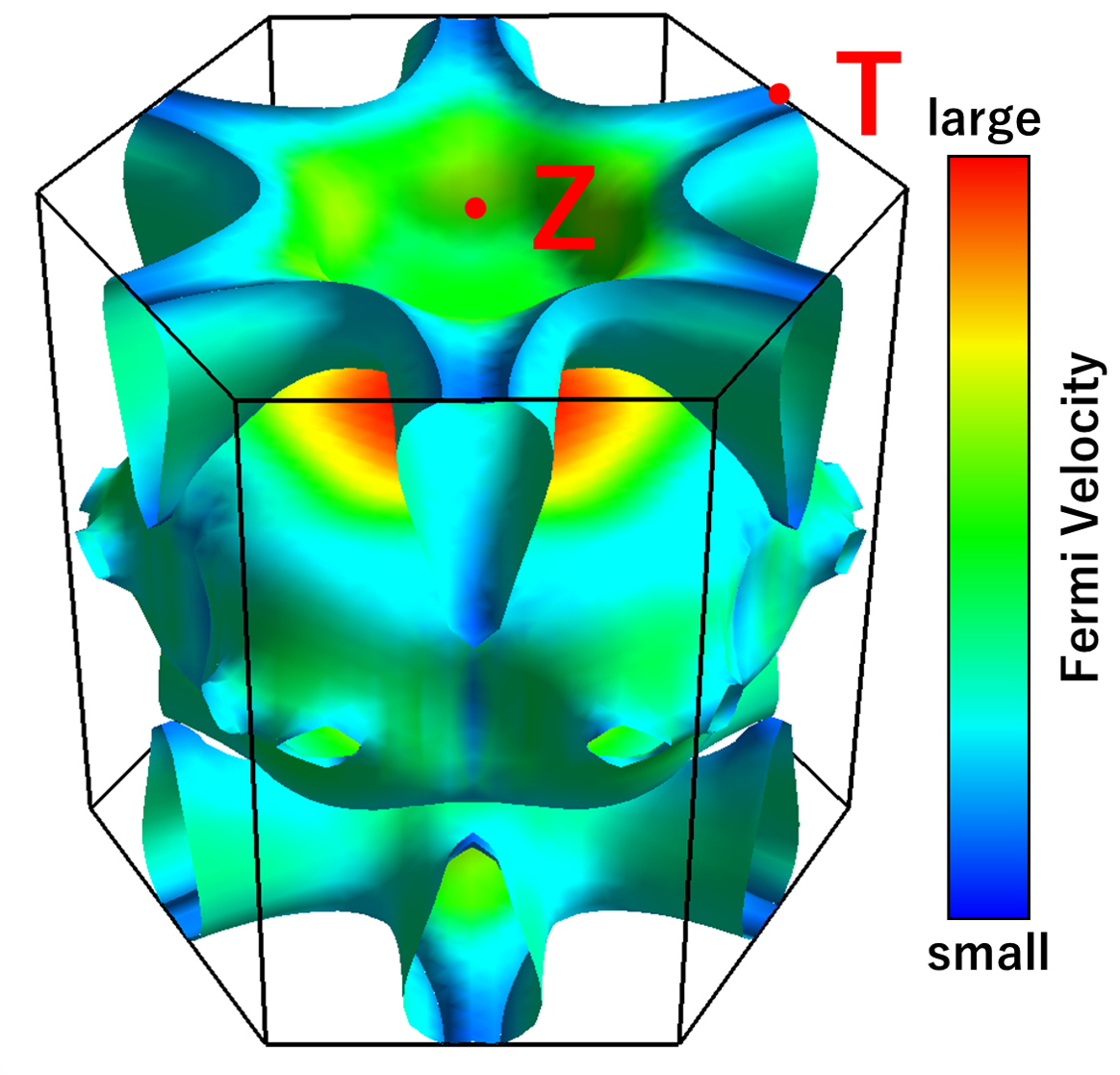}
    \caption{Fermi Surface of Mn$_3$Sn before compression. Fermi surface color map plots the absolute value of Fermi velocity. This figure is created using FermiSurfer~\cite{Kawamurafermisurf}.
    }
    \label{fig:Fermisurf}
\end{figure}

To elucidate the electronic-structure origin of the piezomagnetic effect in Mn$_3$Sn, we analyzed in detail how the spin distribution in $k$-space evolves under uniaxial compression.
Figure~\ref{fig:BRZ+ST3D}(a) shows Brillouin zone (BRZ) of Mn$_3$Sn, and Figure~\ref{fig:BRZ+ST3D}(b)-(d) presents the distribution of the $x$-component of spin moment, $S_{x}({\bm k})$, in $k$-space. 
Figures~\ref{fig:BRZ+ST3D}(b) and ~\ref{fig:BRZ+ST3D}(c) show that the regions where $S_{x}({\bm k})$ takes negative values expand under compression, consistent with the growth of the total spin moment toward the negative direction.
Figure~\ref{fig:BRZ+ST3D}(d) shows the difference in $S_{x}({\bm k})$ between the compressed and uncompressed states. 
The figure reveals that a weak magnetization is induced almost uniformly across the BRZ, whereas strong magnetization is localized to limited regions; comparizon with the Fermi surface in Fig.~\ref{fig:Fermisurf} shows that these strongly magnetized $k$-space regions correspond to those near the Fermi surface. Quantitative evaluation further reveals that these strongly magnetized regions in $k$-space contribute about 10\% of the total spin magnetization.
Although the uniform background is responsible for the majority of the induced magnetization, the contribution originating from the Fermi surface is qualitatively distinct: it reflects the spin response of itinerant carriers, a mechanism that is intrinsic to antiferromagnetic metals and absent in piezomagnetism in antiferromagnetic insulators. Consequently, this Fermi-surface-induced component is not a minor correction but a key microscopic source of the piezomagnetic response in Mn$_3$Sn.

\begin{figure}
    \centering
    \includegraphics[width=1.0\linewidth]{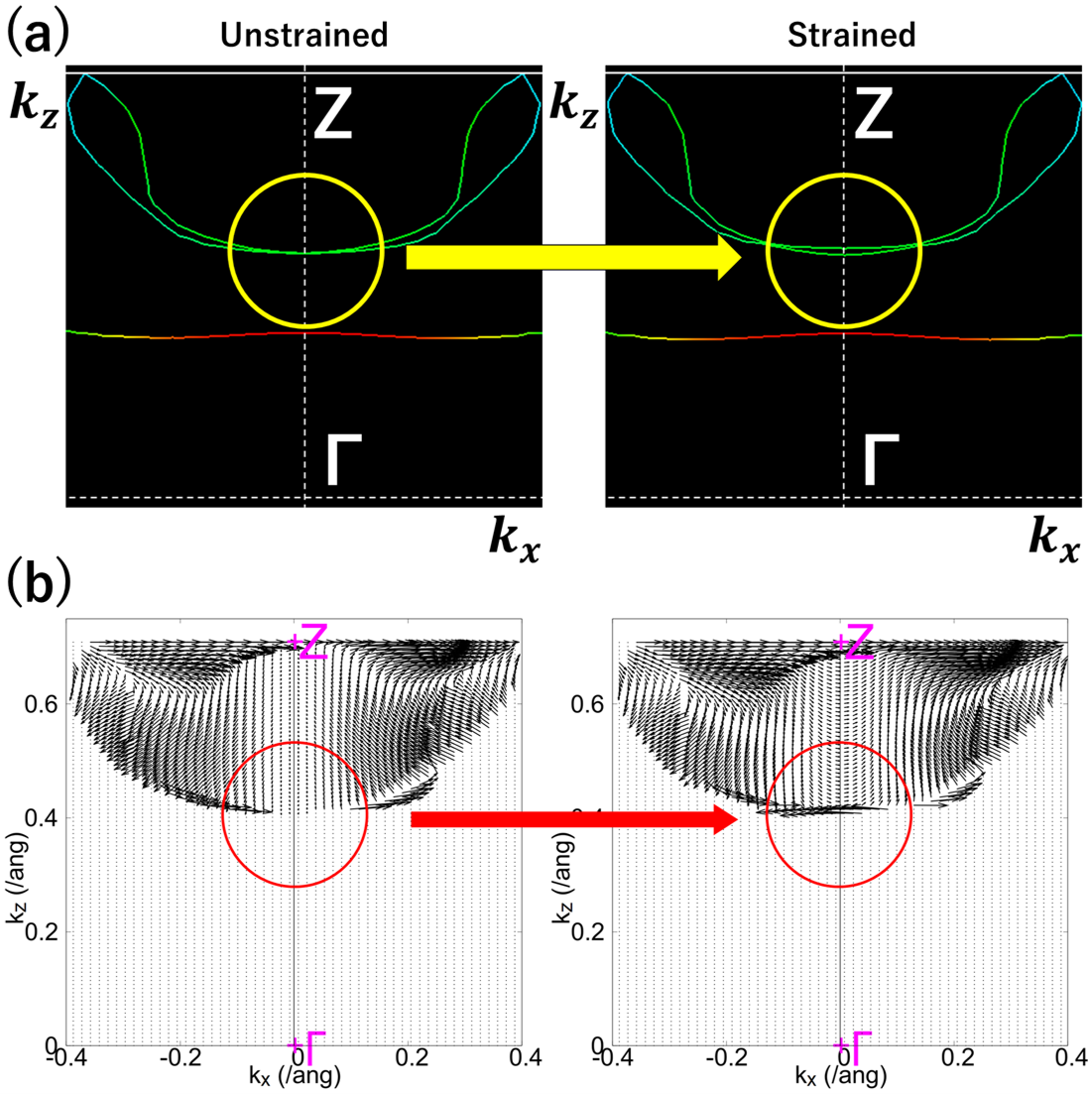}
    \caption{(a) Changes in the Fermi surface due to compression and (b) Spin distribution calculated for two bands forming a split Fermi surface on the $k_xk_z$ plane containing $\Gamma$ and $Z$ points.
    }
    \label{fig:GZ_Fermi_2dST}
\end{figure}

From this point, we focus on a specific plane of the BRZ, particularly the vicinity of the Fermi surface where strong magnetization is induced. Figure~\ref{fig:GZ_Fermi_2dST} presents (a) the Fermi surface before and after compression on the $k_xk_z$ plane containing the $\Gamma$ and $Z$ points, and (b) the corresponding spin distribution ${\bm S}({\bm k})$ in the same $k$-space's plane. In Fig.~\ref{fig:GZ_Fermi_2dST}(b), the spin distribution ${\bm S}({\bm k})$ calculated for two specific bands is illustrated. As highlighted by the yellow circle in Fig.~\ref{fig:GZ_Fermi_2dST}(a), the two pseudo-degenerate Fermi surfaces are split by compression. 
Correspondingly, in the region indicated by the red circle in Fig.~\ref{fig:GZ_Fermi_2dST}(b), compression enhances the spin component oriented along the negative $k_x$-direction. 
These results demonstrate that, in the vicinity of the Fermi surface on the $k_xk_z$ plane containing the $\Gamma$ and $Z$ points, the splitting of pseudo-degenerate bands induced by compression generates magnetization. 

The pseudo-degeneracy of energy bands along the $\Gamma$--$\mathrm{Z}$ axis in the unstrained state can be understood from the electronic states in the SOC-free case. 
 The symmetry of the electronic structure in the magnetically ordered systems with the SOC is characterized by the conventional magnetic space group~\cite{Bradley_book_2010}, whereas that in the SOC-free case is described using the spin crystallographic group~\cite{Brinkman_SSG_1966,LITVIN1974538,Litvin:a14103,Liu_PhysRevX.12.021016,Smejkal_PhysRevX.12.031042,Watanabe_PhysRevB.109.094438,Zhenyu_PhysRevX.14.031037,Xiaobing_PhysRevX.14.031038,Jiang_PhysRevX.14.031039,Schiff_SciPostPhys.18.3.109,Song_PhysRevB.111.134407}. 
 According to the analysis based on the spin space group in Refs.~\cite{Jiang_PhysRevX.14.031039,Song_PhysRevB.111.134407}, 
 two-dimensional irreducible representations  (irreps.) are allowed on the $\Gamma$--$\mathrm{Z}$ line in the unstrained antiferromagnetic state in Mn$_3$Sn without the SOC, while the all irreps. are one-dimensional in the presence of the SOC.  
 The degeneracies in the energy bands belonging to the two-dimensional irreps. in the SOC-free case are lifted by introducing the SOC but the energy splitting is tiny due to the weak SOC on Mn-atom, resulting in the pseudo-degeneracy as shown in the left panel of Fig.~\ref{fig:GZ_Fermi_2dST}(a).  
 The uniaxial pressure along $x$-axis enhances the energy splitting as shown in the right panel of Fig.~\ref{fig:GZ_Fermi_2dST}(b).

\begin{figure}
    \centering
    \includegraphics[width=1.0\linewidth]{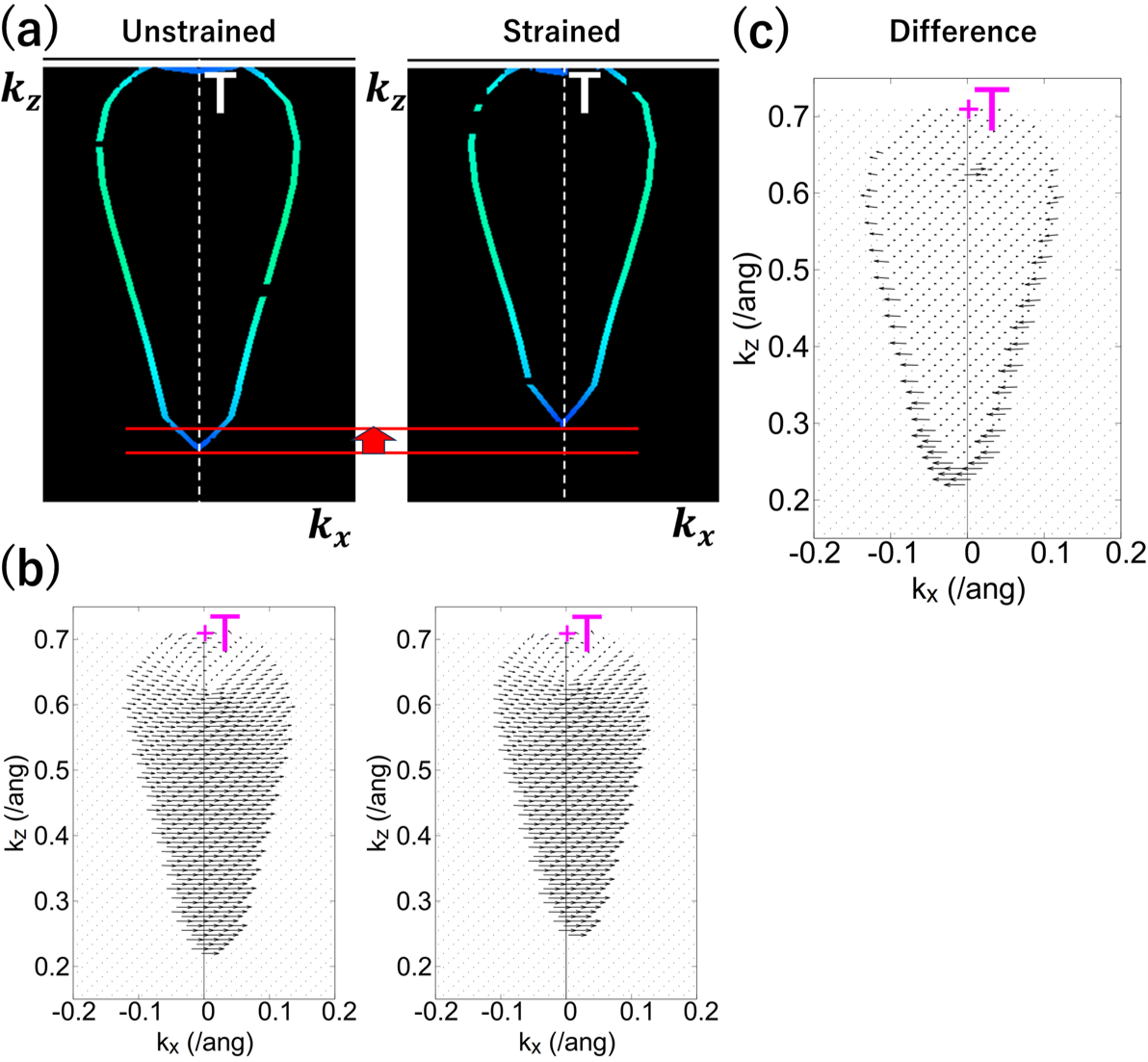}
    \caption{(a) Changes in the Fermi surface due to compression and (b) Spin distribution calculated for the band forming the Fermi surface on the $k_xk_z$ plane containing $Y$ and $T$ points, and (c) its change in the spin distribution as Fig.~\ref{fig:BRZ+ST3D}(d). 
    }
    \label{fig:XY_Fermi+2dST}
\end{figure}

\begin{figure}
    \centering
    \includegraphics[width=1.0\linewidth]{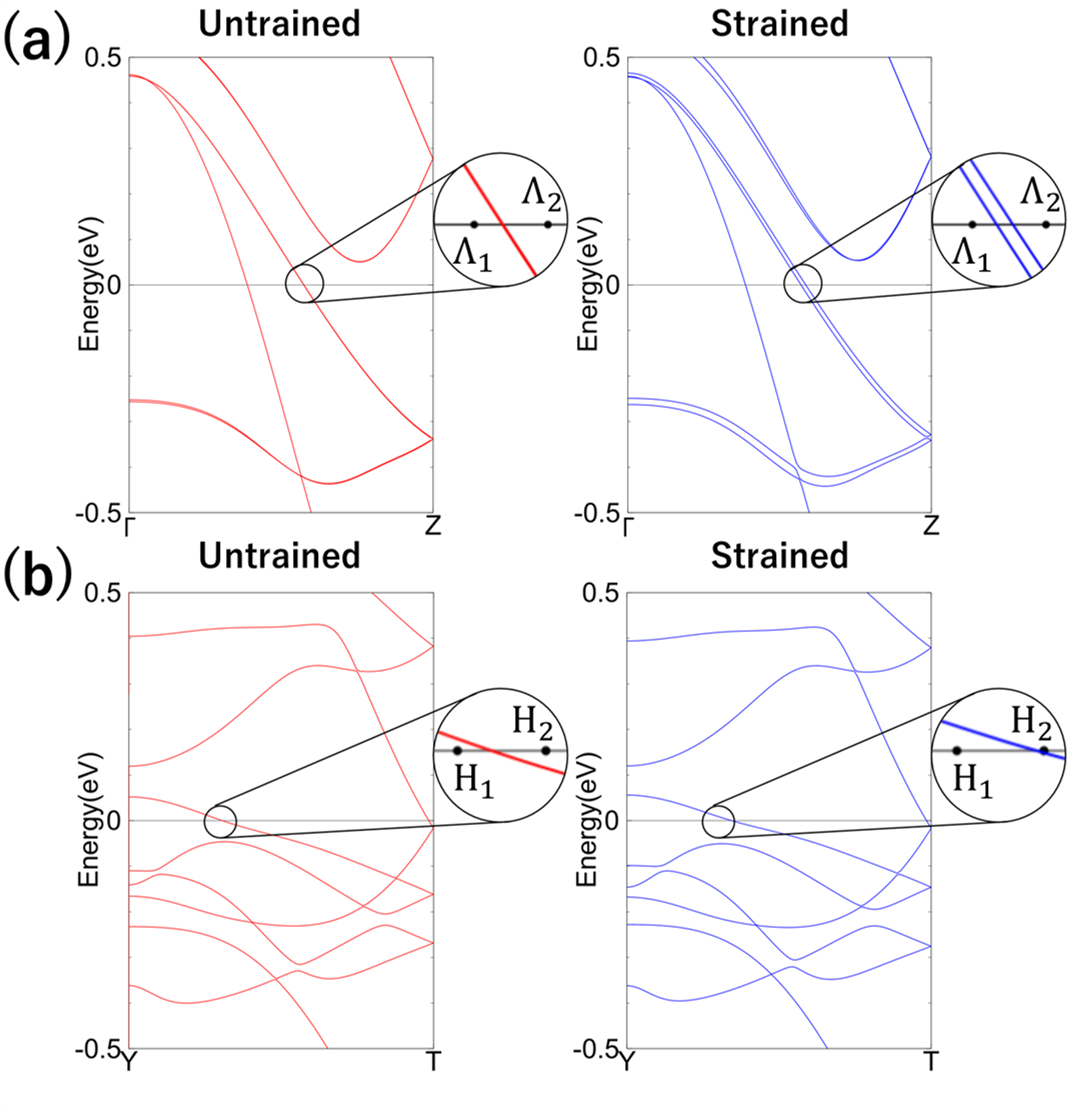}
    \caption{Energy bands along (a) $\Gamma$-$Z$ line and
(b) $Y$-$T$ line of BRZ. The insets are close-up of the energy bands near the Fermi level. 
}
    \label{fig:bandGZ-XY}
\end{figure}

Next, we analyze the planes parallel to the $k_xk_z$ plane including the $Y$ and $T$ points. Figure~\ref{fig:XY_Fermi+2dST} presents: (a) the Fermi surfaces before and after compression; (b) the corresponding spin distribution of $S_{x}({\bm k})$; and (c) the change in the distribution of $S_{x}({\bm k})$. 
Figure~\ref{fig:XY_Fermi+2dST}(a) shows that compression leads to a shrinkage of the Fermi surface as indicated by the red reference lines, reducing the occupied region below the Fermi surface. This shrinkage of the Fermi surface reduces the magnetized region in the BRZ plane.
The plot of the difference before and after compression in Fig.~\ref{fig:XY_Fermi+2dST}(c) further reveals an enhancement of spin oriented in the negative $k_x$ direction within the reduced region. 
Therefore, on planes parallel to the $k_xk_z$ plane containing the $Y$ and $T$ points, this Fermi surface shrinkage induced by compression contributes to the development of spin polarization.
 Figure~\ref{fig:bandGZ-XY} shows the band structures along the $\Gamma$–$Z$ and $Y$–$T$ lines.
Figure~\ref{fig:bandGZ-XY}(a) presents the band splitting corresponding to the Fermi surface splitting in Fig.~\ref{fig:GZ_Fermi_2dST}(a).
Also, Fig.~\ref{fig:bandGZ-XY}(b) exhibits a shift of the energy band toward the positive $k_z$ direction, corresponding to the Fermi surface shift shown in Fig.~\ref{fig:XY_Fermi+2dST}.

\subsection{Relationship between $k$-space and real-space pictures} 
\label{Sec:kxky ky=0}

\begin{figure}
    \centering
    \includegraphics[width=1.0\linewidth]{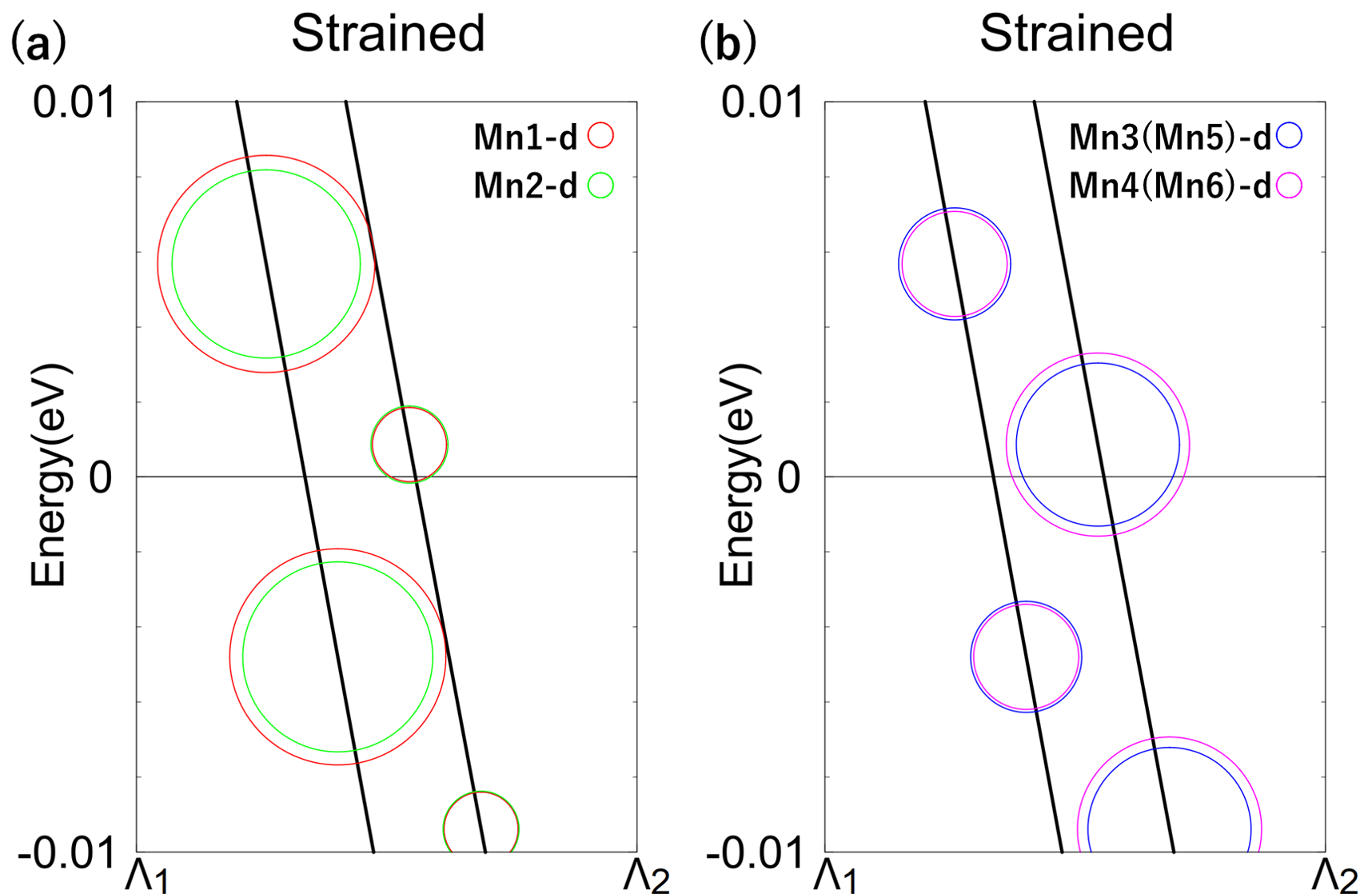}
    \caption{Band structure split of $\Lambda_1$–$\Lambda_2$ line by compression, with orbital decomposition (a) by Mn1-2 atoms $d$-orbitals, (b) by Mn3-6 atoms $d$-orbitals.}
    \label{fig:band_st_Mn-d}
\end{figure}

Here, we discuss the relationship between the $k$-space and real space pictures of the piezomagnetic effect in Mn$_3$Sn, as discussed thus far. Figure~\ref{fig:band_st_Mn-d} shows the energy bands split by compression, corresponding to those shown in the inset of the right panel of Fig.~\ref{fig:bandGZ-XY}(a). In that figure, the magnitude of the contribution from each Mn atom’s $d$-orbitals is represented by the size of differently colored circles placed on the calculated points. Since the contributions from non-Mn $d$-orbitals are negligible near the Fermi level of these bands, they are not shown in the figure.

Figure~\ref{fig:band_st_Mn-d}(a) shows the $d$-orbital contributions from Mn1 and Mn2 and~\ref{fig:band_st_Mn-d}(b) shows those from Mn3–Mn6.
In the $d$-orbital components, the contributions of Mn3 and Mn5, as well as those of Mn4 and Mn6, are identical.
Figures~\ref{fig:band_st_Mn-d}(a) and ~\ref{fig:band_st_Mn-d}(b) indicate that, in one band, the $d$-orbitals of Mn1 and Mn2 constitute the major components, whereas in the other band, the dominant components are the $d$-orbitals of Mn3-Mn6.
The classification of these two bands corresponds to that in Fig.~\ref{fig:un-st_AFM1}, which distinguishes atoms Mn1 and Mn2 whose spin moments are parallel to $k_x$ axis and remain unrotated under compression from atoms Mn3-Mn6 whose spin moments are non-parallel to $k_x$ and rotate under compression. 

Next, we characterize the band shifts induced by compression in terms of the change in the Fermi wave-number $k_F$, defined as the wave-number at which each band crosses the Fermi level.
The compression induces only a $\qty{0.0004}{\per\angstrom}$ change in the $\Gamma$–$Z$ path length, which is negligible compared to the Fermi wave-number shifts that characterize the band displacement discussed below.

We find that the Mn3–Mn6–dominated band exhibits a positive shift $\Delta k_F = +\qty{0.0078}{\per\angstrom}$ along $k_z$, whereas the Mn1–Mn2–dominated band shifts negatively by $\Delta k_F = \qty{-0.0027}{\per\angstrom}$.
These opposite changes in $k_F$ directly account for the band splitting in Fig.~\ref{fig:bandGZ-XY}(a) and the corresponding Fermi-surface separation in Fig.~\ref{fig:GZ_Fermi_2dST}(a).

\begin{figure}
    \centering
    \includegraphics[width=1.0\linewidth]{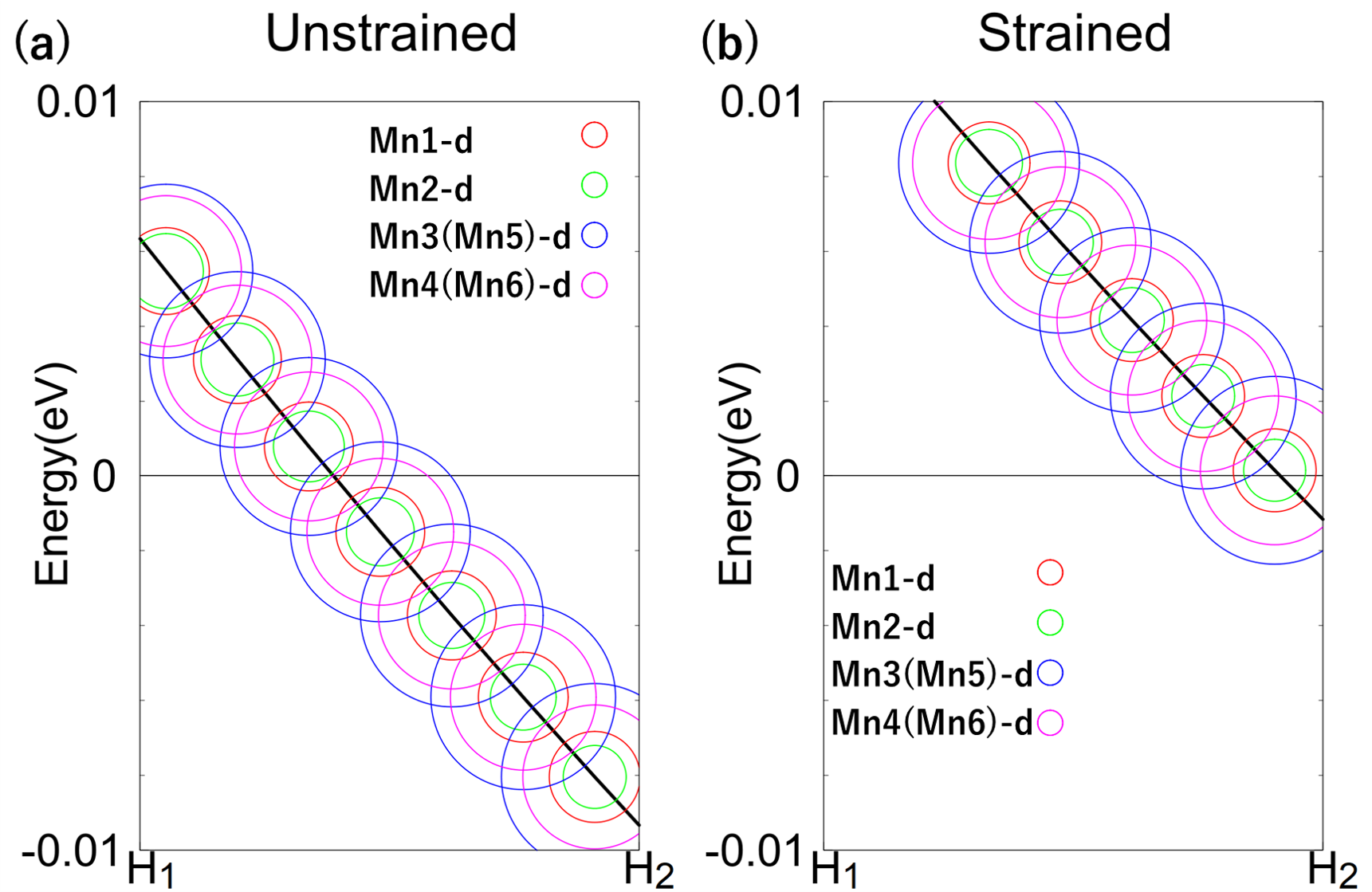}
    \caption{Band structure of $H_1$-$H_2$ line with orbital decomposition (a) before compression (b) after compression.}
    \label{fig:XY_weight_band}
\end{figure}

Along the $H_1$–$H_2$ direction, the bands near the Fermi level are mainly composed of the $d$ orbitals of Mn3–Mn6 (Fig.~\ref{fig:XY_weight_band}).
Their Fermi wave-number, defined relative to the $Y$ point, increases by $\Delta k_F = +\qty{0.0257}{\per\angstrom}$, which correlates with the spin rotation induced on these Mn sites under compression.

\begin{figure}
    \centering
    \includegraphics[width=1.0\linewidth]{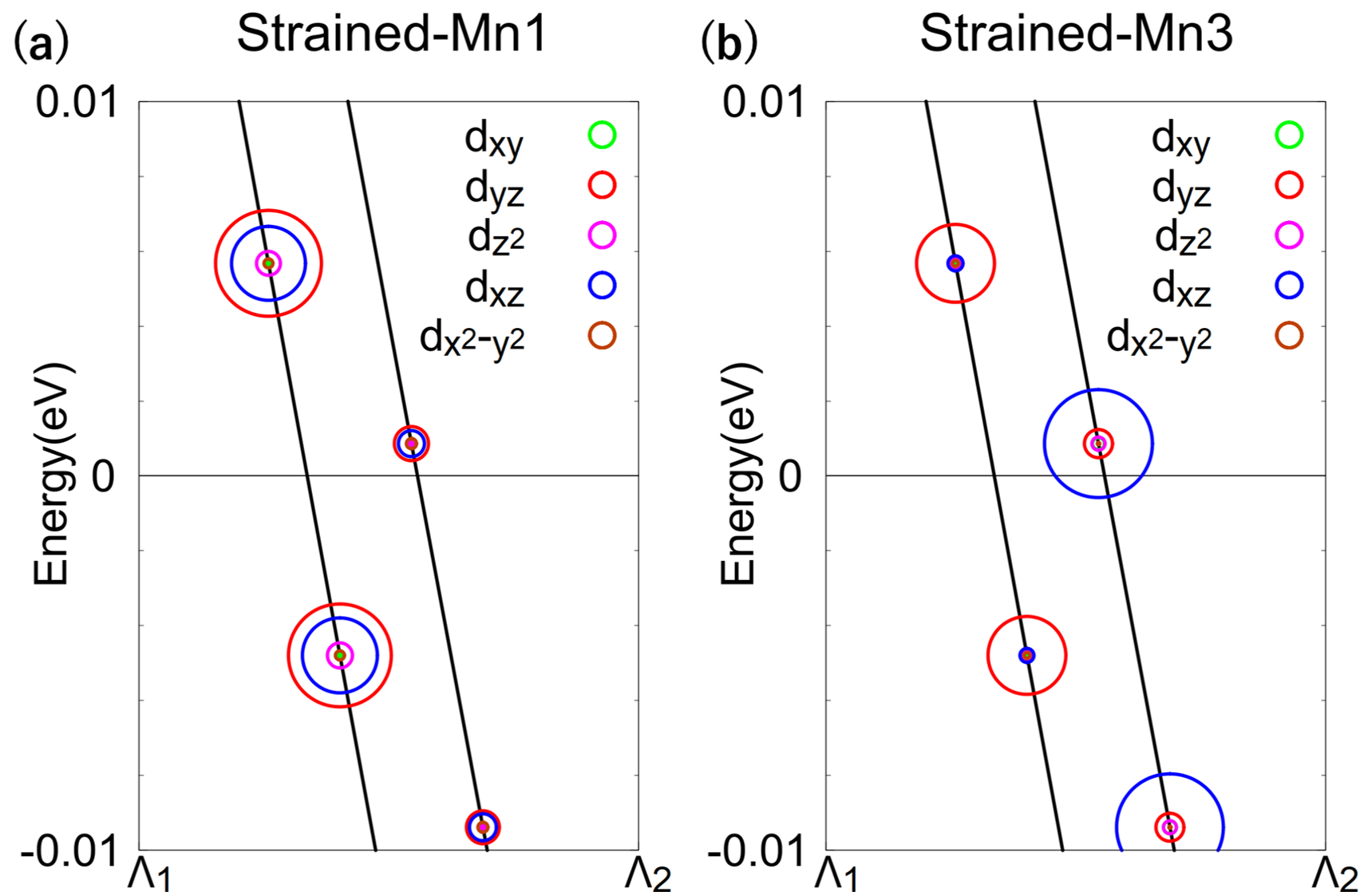}
    \caption{Band structure split of $\Lambda_1$–$\Lambda_2$ line by compression, with orbital decomposition by (a) Mn1 atom and (a) Mn3 atom $d_{xy}, d_{yz}, d_{z^2}, d_{xz}, d_{x^2-y^2}$-orbitals.}
    \label{fig:band_Mn1-3}
\end{figure}

Figure~\ref{fig:band_Mn1-3} presents the decomposition of the $d$-orbital weights into the five components ($d_{xy}$, $d_{yz}$, $d_{z^2}$, $d_{xz}$, and $d_{x^2-y^2}$). Panel (a) shows the orbital contributions for Mn1, while panel (b) shows those for Mn3. Since Mn2 exhibits similar $d$-orbital characteristics as Mn1, and Mn4–Mn6 behave similarly to Mn3, these two panels are representative of the respective groups of Mn atoms.
In panel (a), the left-side band—dominated by the $d$-orbital contributions from Mn1 and Mn2—displays particularly strong weights from the $d_{xz}$ and $d_{yz}$ orbitals. In contrast, panel (b) shows that the right-side band, originating mainly from the $d$-orbitals of Mn3–Mn6 and exhibiting a notably large $d_{xz}$ contribution, shifts toward higher energy under compression.

\begin{figure}
    \centering
    \includegraphics[width=1.0\linewidth]{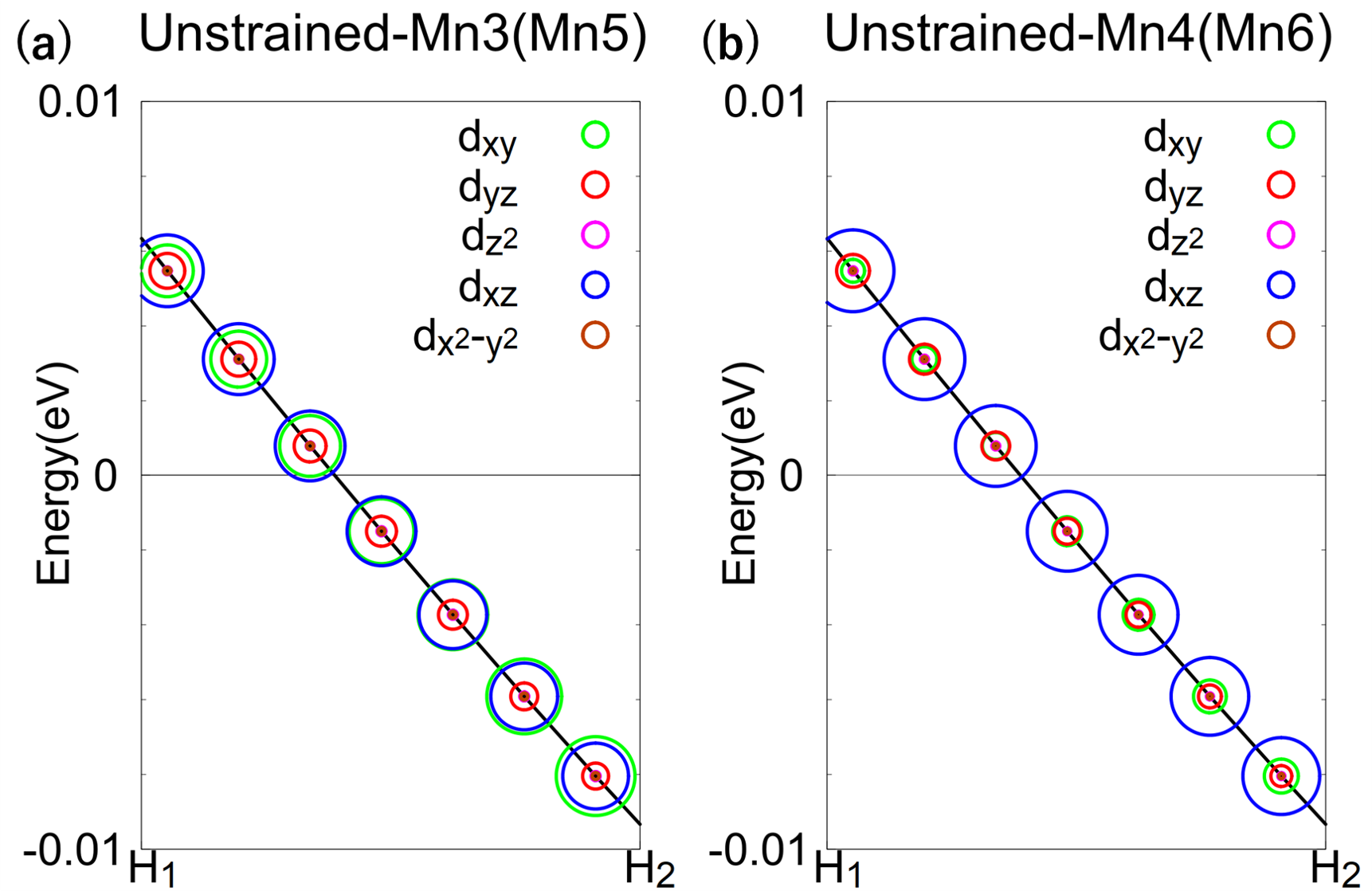}
    \caption{Band structure of $H_1$-$H_2$ line before compression with orbital decomposition by (a) Mn3(Mn5) atoms and (b) Mn4(Mn6) atoms $d_{xy}, d_{yz}, d_{z^2}, d_{xz}, d_{x^2-y^2}$-orbitals.}
    \label{fig:XY_Mn3-Mn6}
\end{figure}

Figure~\ref{fig:XY_Mn3-Mn6} shows the decomposition of Fig.~\ref{fig:XY_weight_band}(a) into the $d_{xy}, d_{yz}, d_{z^2}, d_{xz}, d_{x^2-y^2}$-orbitals on Mn3 and Mn4 atoms. Mn5 and Mn6 have the the similar orbital character with Mn3 and Mn4, shown in Fig.~\ref{fig:XY_Mn3-Mn6}(a) and ~\ref{fig:XY_Mn3-Mn6}(b),  respectively.
The difference in orbital character between the Mn3 (Mn5) atoms and the Mn4 (Mn6) atoms arises from the phase of the single-electron wave function with finite $k_y$, which affects atoms located at different $y$ coordinates in real space.
Figure~\ref{fig:XY_Mn3-Mn6}(a) shows the orbital weights of Mn3 (Mn5), and Fig.~\ref{fig:XY_Mn3-Mn6}(b) shows those of Mn4 (Mn6). As in Fig.~\ref{fig:band_Mn1-3}, the $d_{xz}$-orbital contribution is dominant for both Mn3 (Mn5) and Mn4 (Mn6), and Mn3 (Mn5) also exhibits a significant contribution from the $d_{xy}$ orbital.

These results establish a direct correspondence between spin reorientation in real space under compression and the modifications of the electronic structure in $k$-space, thereby providing a microscopic understanding of the magnetization mechanism.

\section{CONCLUSION}
We investigated the microscopic mechanism of magnetization induction in the piezomagnetic effect of Mn$_3$Sn through a detailed electronic structure analysis. Our results demonstrate a direct correspondence between spin reorientation in real space and band shifts in $k$-space. In particular, compression induces a positive $k_z$-direction shift in bands dominated by the $d_{xz}$ and $d_{xy}$-orbitals of Mn3-Mn6 atoms, whose spin moments rotate under strain. In contrast, Mn1 and Mn2 atoms exhibit no spin moment rotation, and their associated bands show only a minor negative $k_z$-direction shift of the energy band. 
These findings establish a clear connection between real-space spin dynamics and $k$-space electronic structure, providing a microscopic understanding of stress-induced magnetization in Mn$_3$Sn. This framework offers new insight into the mechanism of piezomagnetism and may serve as a foundation for future studies on pressure-induced magnetic phenomena and their potential applications in spintronics.

\begin{acknowledgements}
 We are grateful to K. Tanno for the technical supports. We also thank I. Terada for helpful comments and discussions. This research is supported by JSPS KAKENHI Grants Numbers JP23H01130, JP24K00581, JP24K00588, JP25K00947, JP25K21684, and JP23K20824.
\end{acknowledgements}

\bibliography{Mn3SnPiezo}

\begin{thebibliography}{52}%
\makeatletter
\providecommand \@ifxundefined [1]{%
 \@ifx{#1\undefined}
}%
\providecommand \@ifnum [1]{%
 \ifnum #1\expandafter \@firstoftwo
 \else \expandafter \@secondoftwo
 \fi
}%
\providecommand \@ifx [1]{%
 \ifx #1\expandafter \@firstoftwo
 \else \expandafter \@secondoftwo
 \fi
}%
\providecommand \natexlab [1]{#1}%
\providecommand \enquote  [1]{``#1''}%
\providecommand \bibnamefont  [1]{#1}%
\providecommand \bibfnamefont [1]{#1}%
\providecommand \citenamefont [1]{#1}%
\providecommand \href@noop [0]{\@secondoftwo}%
\providecommand \href [0]{\begingroup \@sanitize@url \@href}%
\providecommand \@href[1]{\@@startlink{#1}\@@href}%
\providecommand \@@href[1]{\endgroup#1\@@endlink}%
\providecommand \@sanitize@url [0]{\catcode `\\12\catcode `\$12\catcode `\&12\catcode `\#12\catcode `\^12\catcode `\_12\catcode `\%12\relax}%
\providecommand \@@startlink[1]{}%
\providecommand \@@endlink[0]{}%
\providecommand \url  [0]{\begingroup\@sanitize@url \@url }%
\providecommand \@url [1]{\endgroup\@href {#1}{\urlprefix }}%
\providecommand \urlprefix  [0]{URL }%
\providecommand \Eprint [0]{\href }%
\providecommand \doibase [0]{https://doi.org/}%
\providecommand \selectlanguage [0]{\@gobble}%
\providecommand \bibinfo  [0]{\@secondoftwo}%
\providecommand \bibfield  [0]{\@secondoftwo}%
\providecommand \translation [1]{[#1]}%
\providecommand \BibitemOpen [0]{}%
\providecommand \bibitemStop [0]{}%
\providecommand \bibitemNoStop [0]{.\EOS\space}%
\providecommand \EOS [0]{\spacefactor3000\relax}%
\providecommand \BibitemShut  [1]{\csname bibitem#1\endcsname}%
\let\auto@bib@innerbib\@empty
\bibitem [{\citenamefont {Tomiyoshi}\ and\ \citenamefont {Yamaguchi}(1982)}]{Tomiyoshi1982}%
  \BibitemOpen
  \bibfield  {author} {\bibinfo {author} {\bibfnamefont {S.}~\bibnamefont {Tomiyoshi}}\ and\ \bibinfo {author} {\bibfnamefont {Y.}~\bibnamefont {Yamaguchi}},\ }\bibfield  {title} {\bibinfo {title} {{Magnetic Structure and Weak Ferromagnetism of Mn$_{3}$Sn Studied by Polarized Neutron Diffraction}},\ }\href {https://doi.org/10.1143/JPSJ.51.2478} {\bibfield  {journal} {\bibinfo  {journal} {J. Phys. Soc. Jpn.}\ }\textbf {\bibinfo {volume} {51}},\ \bibinfo {pages} {2478} (\bibinfo {year} {1982})}\BibitemShut {NoStop}%
\bibitem [{\citenamefont {Brown}\ \emph {et~al.}(1990)\citenamefont {Brown}, \citenamefont {Nunez}, \citenamefont {Tasset}, \citenamefont {Forsyth},\ and\ \citenamefont {Radhakrishna}}]{Brown1990}%
  \BibitemOpen
  \bibfield  {author} {\bibinfo {author} {\bibfnamefont {P.~J.}\ \bibnamefont {Brown}}, \bibinfo {author} {\bibfnamefont {V.}~\bibnamefont {Nunez}}, \bibinfo {author} {\bibfnamefont {F.}~\bibnamefont {Tasset}}, \bibinfo {author} {\bibfnamefont {J.~B.}\ \bibnamefont {Forsyth}},\ and\ \bibinfo {author} {\bibfnamefont {P.}~\bibnamefont {Radhakrishna}},\ }\bibfield  {title} {\bibinfo {title} {{Determination of the magnetic structure of Mn$_3$Sn using generalized neutron polarization analysis}},\ }\href {https://doi.org/10.1088/0953-8984/2/47/015} {\bibfield  {journal} {\bibinfo  {journal} {J. Phys.^^e2^^80^^93Condens. Matter.}\ }\textbf {\bibinfo {volume} {2}},\ \bibinfo {pages} {9409} (\bibinfo {year} {1990})}\BibitemShut {NoStop}%
\bibitem [{\citenamefont {Nakatsuji}\ \emph {et~al.}(2015)\citenamefont {Nakatsuji}, \citenamefont {Kiyohara},\ and\ \citenamefont {Higo}}]{Nakatsuji2015}%
  \BibitemOpen
  \bibfield  {author} {\bibinfo {author} {\bibfnamefont {S.}~\bibnamefont {Nakatsuji}}, \bibinfo {author} {\bibfnamefont {N.}~\bibnamefont {Kiyohara}},\ and\ \bibinfo {author} {\bibfnamefont {T.}~\bibnamefont {Higo}},\ }\bibfield  {title} {\bibinfo {title} {{Large anomalous Hall effect in a non-collinear antiferromagnet at room temperature}},\ }\href {https://doi.org/10.1038/nature15723} {\bibfield  {journal} {\bibinfo  {journal} {Nat.}\ }\textbf {\bibinfo {volume} {527}},\ \bibinfo {pages} {212} (\bibinfo {year} {2015})}\BibitemShut {NoStop}%
\bibitem [{\citenamefont {Suzuki}\ \emph {et~al.}(2017)\citenamefont {Suzuki}, \citenamefont {Koretsune}, \citenamefont {Ochi},\ and\ \citenamefont {Arita}}]{Suzuki2017}%
  \BibitemOpen
  \bibfield  {author} {\bibinfo {author} {\bibfnamefont {M.-T.}\ \bibnamefont {Suzuki}}, \bibinfo {author} {\bibfnamefont {T.}~\bibnamefont {Koretsune}}, \bibinfo {author} {\bibfnamefont {M.}~\bibnamefont {Ochi}},\ and\ \bibinfo {author} {\bibfnamefont {R.}~\bibnamefont {Arita}},\ }\bibfield  {title} {\bibinfo {title} {{Cluster multipole theory for anomalous Hall effect in antiferromagnets}},\ }\href {https://doi.org/10.1103/PhysRevB.95.094406} {\bibfield  {journal} {\bibinfo  {journal} {Phys. Rev. B}\ }\textbf {\bibinfo {volume} {95}},\ \bibinfo {pages} {094406} (\bibinfo {year} {2017})}\BibitemShut {NoStop}%
\bibitem [{\citenamefont {Higo}\ \emph {et~al.}(2018)\citenamefont {Higo}, \citenamefont {Man}, \citenamefont {Gopman}, \citenamefont {Wu}, \citenamefont {Koretsune}, \citenamefont {van~'t Erve}, \citenamefont {Kabanov}, \citenamefont {Rees}, \citenamefont {Li}, \citenamefont {Suzuki}, \citenamefont {Patankar}, \citenamefont {Ikhlas}, \citenamefont {Chien}, \citenamefont {Arita}, \citenamefont {Shull}, \citenamefont {Orenstein},\ and\ \citenamefont {Nakatsuji}}]{Higo2018}%
  \BibitemOpen
  \bibfield  {author} {\bibinfo {author} {\bibfnamefont {T.}~\bibnamefont {Higo}}, \bibinfo {author} {\bibfnamefont {H.}~\bibnamefont {Man}}, \bibinfo {author} {\bibfnamefont {D.~B.}\ \bibnamefont {Gopman}}, \bibinfo {author} {\bibfnamefont {L.}~\bibnamefont {Wu}}, \bibinfo {author} {\bibfnamefont {T.}~\bibnamefont {Koretsune}}, \bibinfo {author} {\bibfnamefont {O.~M.~J.}\ \bibnamefont {van~'t Erve}}, \bibinfo {author} {\bibfnamefont {Y.~P.}\ \bibnamefont {Kabanov}}, \bibinfo {author} {\bibfnamefont {D.}~\bibnamefont {Rees}}, \bibinfo {author} {\bibfnamefont {Y.}~\bibnamefont {Li}}, \bibinfo {author} {\bibfnamefont {M.-T.}\ \bibnamefont {Suzuki}}, \bibinfo {author} {\bibfnamefont {S.}~\bibnamefont {Patankar}}, \bibinfo {author} {\bibfnamefont {M.}~\bibnamefont {Ikhlas}}, \bibinfo {author} {\bibfnamefont {C.~L.}\ \bibnamefont {Chien}}, \bibinfo {author} {\bibfnamefont {R.}~\bibnamefont {Arita}}, \bibinfo {author} {\bibfnamefont {R.~D.}\ \bibnamefont {Shull}}, \bibinfo {author} {\bibfnamefont {J.}~\bibnamefont
  {Orenstein}},\ and\ \bibinfo {author} {\bibfnamefont {S.}~\bibnamefont {Nakatsuji}},\ }\bibfield  {title} {\bibinfo {title} {{Large magneto-optical Kerr effect and imaging of magnetic octupole domains in an antiferromagnetic metal}},\ }\href {https://doi.org/10.1038/s41566-017-0086-z} {\bibfield  {journal} {\bibinfo  {journal} {Nat. Photonics}\ }\textbf {\bibinfo {volume} {12}},\ \bibinfo {pages} {73} (\bibinfo {year} {2018})}\BibitemShut {NoStop}%
\bibitem [{\citenamefont {Ikhlas}\ \emph {et~al.}(2022)\citenamefont {Ikhlas}, \citenamefont {Dasgupta}, \citenamefont {Theuss}, \citenamefont {Higo}, \citenamefont {Kittaka}, \citenamefont {Ramshaw}, \citenamefont {Tchernyshyov}, \citenamefont {Hicks},\ and\ \citenamefont {Nakatsuji}}]{Ikhlas2022}%
  \BibitemOpen
  \bibfield  {author} {\bibinfo {author} {\bibfnamefont {M.}~\bibnamefont {Ikhlas}}, \bibinfo {author} {\bibfnamefont {S.}~\bibnamefont {Dasgupta}}, \bibinfo {author} {\bibfnamefont {F.}~\bibnamefont {Theuss}}, \bibinfo {author} {\bibfnamefont {T.}~\bibnamefont {Higo}}, \bibinfo {author} {\bibfnamefont {S.}~\bibnamefont {Kittaka}}, \bibinfo {author} {\bibfnamefont {B.~J.}\ \bibnamefont {Ramshaw}}, \bibinfo {author} {\bibfnamefont {O.}~\bibnamefont {Tchernyshyov}}, \bibinfo {author} {\bibfnamefont {C.~W.}\ \bibnamefont {Hicks}},\ and\ \bibinfo {author} {\bibfnamefont {S.}~\bibnamefont {Nakatsuji}},\ }\bibfield  {title} {\bibinfo {title} {{Piezomagnetic switching of the anomalous Hall effect in an antiferromagnet at room temperature}},\ }\href {https://doi.org/10.1038/s41567-022-01640-4} {\bibfield  {journal} {\bibinfo  {journal} {Nat. Phys.}\ }\textbf {\bibinfo {volume} {18}},\ \bibinfo {pages} {1086} (\bibinfo {year} {2022})}\BibitemShut {NoStop}%
\bibitem [{\citenamefont {Dzialoshinskii}(1958)}]{Dzialoshinskii1958}%
  \BibitemOpen
  \bibfield  {author} {\bibinfo {author} {\bibfnamefont {I.~E.}\ \bibnamefont {Dzialoshinskii}},\ }\bibfield  {title} {\bibinfo {title} {{The problem of piezomagnetism}},\ }\href@noop {} {\bibfield  {journal} {\bibinfo  {journal} {Soviet Physics JETP}\ }\textbf {\bibinfo {volume} {6}},\ \bibinfo {pages} {621} (\bibinfo {year} {1958})}\BibitemShut {NoStop}%
\bibitem [{\citenamefont {Moriya}(1959)}]{Moriya1959}%
  \BibitemOpen
  \bibfield  {author} {\bibinfo {author} {\bibfnamefont {T.}~\bibnamefont {Moriya}},\ }\bibfield  {title} {\bibinfo {title} {{Piezomagnetism in CoF$_2$}},\ }\href {https://doi.org/10.1016/0022-3697(59)90043-5} {\bibfield  {journal} {\bibinfo  {journal} {J. Phys. Chem. Solids}\ }\textbf {\bibinfo {volume} {11}},\ \bibinfo {pages} {73} (\bibinfo {year} {1959})}\BibitemShut {NoStop}%
\bibitem [{\citenamefont {Borovik-Romanov}(1994)}]{BorovikRomanov1994}%
  \BibitemOpen
  \bibfield  {author} {\bibinfo {author} {\bibfnamefont {A.~S.}\ \bibnamefont {Borovik-Romanov}},\ }\bibfield  {title} {\bibinfo {title} {{Piezomagnetism, linear magnetostriction and magnetooptic effect}},\ }\href {https://doi.org/10.1080/00150199408245071} {\bibfield  {journal} {\bibinfo  {journal} {Ferroelectrics}\ }\textbf {\bibinfo {volume} {162}},\ \bibinfo {pages} {153} (\bibinfo {year} {1994})}\BibitemShut {NoStop}%
\bibitem [{\citenamefont {Borovik-Romanov}(1960)}]{BorovikRomanov1960}%
  \BibitemOpen
  \bibfield  {author} {\bibinfo {author} {\bibfnamefont {A.~S.}\ \bibnamefont {Borovik-Romanov}},\ }\bibfield  {title} {\bibinfo {title} {{Piezomagnetism in the antiferromagnetic fluorides of cobalt and manganese}},\ }\href@noop {} {\bibfield  {journal} {\bibinfo  {journal} {Sov. Phys. JETP}\ }\textbf {\bibinfo {volume} {11}},\ \bibinfo {pages} {786} (\bibinfo {year} {1960})}\BibitemShut {NoStop}%
\bibitem [{\citenamefont {Stout}\ and\ \citenamefont {Catalano}(1953)}]{Stout_1953}%
  \BibitemOpen
  \bibfield  {author} {\bibinfo {author} {\bibfnamefont {J.~W.}\ \bibnamefont {Stout}}\ and\ \bibinfo {author} {\bibfnamefont {E.}~\bibnamefont {Catalano}},\ }\bibfield  {title} {\bibinfo {title} {{Thermal Anomalies Associated with the Antiferromagnetic Ordering of Fe${\mathrm{F}}_{2}$, Co${\mathrm{F}}_{2}$, and Ni${\mathrm{F}}_{2}$}},\ }\href {https://doi.org/10.1103/PhysRev.92.1575} {\bibfield  {journal} {\bibinfo  {journal} {Phys. Rev.}\ }\textbf {\bibinfo {volume} {92}},\ \bibinfo {pages} {1575} (\bibinfo {year} {1953})}\BibitemShut {NoStop}%
\bibitem [{\citenamefont {Stout}\ and\ \citenamefont {Adams}(1942)}]{Stout_1942}%
  \BibitemOpen
  \bibfield  {author} {\bibinfo {author} {\bibfnamefont {J.~W.}\ \bibnamefont {Stout}}\ and\ \bibinfo {author} {\bibfnamefont {H.~E.}\ \bibnamefont {Adams}},\ }\bibfield  {title} {\bibinfo {title} {{Magnetism and the Third Law of Thermodynamics. The Heat Capacity of Manganous Fluoride from 13 to 320°K.}},\ }\href {https://doi.org/10.1021/ja01259a013} {\bibfield  {journal} {\bibinfo  {journal} {Journal of the American Chemical Society}\ }\textbf {\bibinfo {volume} {64}},\ \bibinfo {pages} {1535} (\bibinfo {year} {1942})}\BibitemShut {NoStop}%
\bibitem [{\citenamefont {Andratskii}\ and\ \citenamefont {Borovik-Romanov}(1966)}]{Andratskii_1966}%
  \BibitemOpen
  \bibfield  {author} {\bibinfo {author} {\bibfnamefont {V.~P.}\ \bibnamefont {Andratskii}}\ and\ \bibinfo {author} {\bibfnamefont {A.~S.}\ \bibnamefont {Borovik-Romanov}},\ }\bibfield  {title} {\bibinfo {title} {{PIEZOMAGNETIC EFFECT IN $\alpha$-{Fe}$_2${O}$_3$}},\ }\href@noop {} {\bibfield  {journal} {\bibinfo  {journal} {Sov. Phys. JETP}\ }\textbf {\bibinfo {volume} {24}},\ \bibinfo {pages} {687} (\bibinfo {year} {1966})}\BibitemShut {NoStop}%
\bibitem [{\citenamefont {Phillips}\ \emph {et~al.}(1967)\citenamefont {Phillips}, \citenamefont {Townsend},\ and\ \citenamefont {White}}]{Phillips1967}%
  \BibitemOpen
  \bibfield  {author} {\bibinfo {author} {\bibfnamefont {T.~G.}\ \bibnamefont {Phillips}}, \bibinfo {author} {\bibfnamefont {R.~L.}\ \bibnamefont {Townsend}},\ and\ \bibinfo {author} {\bibfnamefont {R.~L.}\ \bibnamefont {White}},\ }\bibfield  {title} {\bibinfo {title} {{Piezomagnetism of $\ensuremath{\alpha}$-$\mathrm{{Fe}}_{2}$$\mathrm{{O}}_{3}$ and the Magnetoelastic Tensor of $\mathrm{{Fe}}^{3+}$ in $\mathrm{{Al}}_{2}$$\mathrm{{O}}_{3}$}},\ }\href {https://doi.org/10.1103/PhysRev.162.382} {\bibfield  {journal} {\bibinfo  {journal} {Phys. Rev.}\ }\textbf {\bibinfo {volume} {162}},\ \bibinfo {pages} {382} (\bibinfo {year} {1967})}\BibitemShut {NoStop}%
\bibitem [{\citenamefont {Kadomtseva}\ \emph {et~al.}(1981{\natexlab{a}})\citenamefont {Kadomtseva}, \citenamefont {Agafonov}, \citenamefont {Milov}, \citenamefont {Moskvin},\ and\ \citenamefont {Semenov}}]{Kadomtseva_JETPLett1981}%
  \BibitemOpen
  \bibfield  {author} {\bibinfo {author} {\bibfnamefont {A.~M.}\ \bibnamefont {Kadomtseva}}, \bibinfo {author} {\bibfnamefont {A.~P.}\ \bibnamefont {Agafonov}}, \bibinfo {author} {\bibfnamefont {V.~N.}\ \bibnamefont {Milov}}, \bibinfo {author} {\bibfnamefont {A.~S.}\ \bibnamefont {Moskvin}},\ and\ \bibinfo {author} {\bibfnamefont {V.~A.}\ \bibnamefont {Semenov}},\ }\bibfield  {title} {\bibinfo {title} {{Direct observation of a symmetry change induced in orthoferrite crystals by an external magnetic field}},\ }\href@noop {} {\bibfield  {journal} {\bibinfo  {journal} {JETP Lett.}\ }\textbf {\bibinfo {volume} {33}},\ \bibinfo {pages} {383} (\bibinfo {year} {1981}{\natexlab{a}})}\BibitemShut {NoStop}%
\bibitem [{\citenamefont {Kadomtseva}\ \emph {et~al.}(1981{\natexlab{b}})\citenamefont {Kadomtseva}, \citenamefont {Agafonov}, \citenamefont {Lukina}, \citenamefont {Milov}, \citenamefont {Moskvin}, \citenamefont {Semenov},\ and\ \citenamefont {Sinitsyn}}]{Kadomtseva_JETP1981}%
  \BibitemOpen
  \bibfield  {author} {\bibinfo {author} {\bibfnamefont {A.~M.}\ \bibnamefont {Kadomtseva}}, \bibinfo {author} {\bibfnamefont {A.~P.}\ \bibnamefont {Agafonov}}, \bibinfo {author} {\bibfnamefont {M.~M.}\ \bibnamefont {Lukina}}, \bibinfo {author} {\bibfnamefont {V.}~\bibnamefont {Milov}}, \bibinfo {author} {\bibfnamefont {A.}~\bibnamefont {Moskvin}}, \bibinfo {author} {\bibfnamefont {V.}~\bibnamefont {Semenov}},\ and\ \bibinfo {author} {\bibfnamefont {E.}~\bibnamefont {Sinitsyn}},\ }\bibfield  {title} {\bibinfo {title} {{Nature of magnetic anisotropy and magnetostriction of orthoferrites and orthochromites}},\ }\href@noop {} {\bibfield  {journal} {\bibinfo  {journal} {Sov. Phys. JETP}\ }\textbf {\bibinfo {volume} {54}},\ \bibinfo {pages} {374} (\bibinfo {year} {1981}{\natexlab{b}})}\BibitemShut {NoStop}%
\bibitem [{\citenamefont {Zvezdin}\ \emph {et~al.}(1985)\citenamefont {Zvezdin}, \citenamefont {Zorin}, \citenamefont {Kadomtseva}, \citenamefont {Krynetskii}, \citenamefont {Moskvin},\ and\ \citenamefont {Mukhin}}]{Zvezdin_JETP1985}%
  \BibitemOpen
  \bibfield  {author} {\bibinfo {author} {\bibfnamefont {A.~K.}\ \bibnamefont {Zvezdin}}, \bibinfo {author} {\bibfnamefont {I.~A.}\ \bibnamefont {Zorin}}, \bibinfo {author} {\bibfnamefont {A.~M.}\ \bibnamefont {Kadomtseva}}, \bibinfo {author} {\bibfnamefont {I.~B.}\ \bibnamefont {Krynetskii}}, \bibinfo {author} {\bibfnamefont {A.~S.}\ \bibnamefont {Moskvin}},\ and\ \bibinfo {author} {\bibfnamefont {A.~A.}\ \bibnamefont {Mukhin}},\ }\bibfield  {title} {\bibinfo {title} {{Linear magnetostriction and antiferromagnetic domain structure in dysprosium orthoferrite}},\ }\href@noop {} {\bibfield  {journal} {\bibinfo  {journal} {Sov. Phys. JETP}\ }\textbf {\bibinfo {volume} {61}},\ \bibinfo {pages} {645} (\bibinfo {year} {1985})}\BibitemShut {NoStop}%
\bibitem [{\citenamefont {Aoyama}\ and\ \citenamefont {Ohgushi}(2024)}]{Aoyama_PhysRevMaterials.8.L041402}%
  \BibitemOpen
  \bibfield  {author} {\bibinfo {author} {\bibfnamefont {T.}~\bibnamefont {Aoyama}}\ and\ \bibinfo {author} {\bibfnamefont {K.}~\bibnamefont {Ohgushi}},\ }\bibfield  {title} {\bibinfo {title} {{Piezomagnetic properties in altermagnetic MnTe}},\ }\href {https://doi.org/10.1103/PhysRevMaterials.8.L041402} {\bibfield  {journal} {\bibinfo  {journal} {Phys. Rev. Mater.}\ }\textbf {\bibinfo {volume} {8}},\ \bibinfo {pages} {L041402} (\bibinfo {year} {2024})}\BibitemShut {NoStop}%
\bibitem [{\citenamefont {Jaime}\ \emph {et~al.}(2017)\citenamefont {Jaime}, \citenamefont {Saul}, \citenamefont {Salamon}, \citenamefont {Zapf}, \citenamefont {Harrison}, \citenamefont {Durakiewicz}, \citenamefont {Lashley}, \citenamefont {Andersson}, \citenamefont {Stanek}, \citenamefont {Smith},\ and\ \citenamefont {Gofryk}}]{Jaime2017}%
  \BibitemOpen
  \bibfield  {author} {\bibinfo {author} {\bibfnamefont {M.}~\bibnamefont {Jaime}}, \bibinfo {author} {\bibfnamefont {A.}~\bibnamefont {Saul}}, \bibinfo {author} {\bibfnamefont {M.}~\bibnamefont {Salamon}}, \bibinfo {author} {\bibfnamefont {V.~S.}\ \bibnamefont {Zapf}}, \bibinfo {author} {\bibfnamefont {N.}~\bibnamefont {Harrison}}, \bibinfo {author} {\bibfnamefont {T.}~\bibnamefont {Durakiewicz}}, \bibinfo {author} {\bibfnamefont {J.~C.}\ \bibnamefont {Lashley}}, \bibinfo {author} {\bibfnamefont {D.~A.}\ \bibnamefont {Andersson}}, \bibinfo {author} {\bibfnamefont {C.~R.}\ \bibnamefont {Stanek}}, \bibinfo {author} {\bibfnamefont {J.~L.}\ \bibnamefont {Smith}},\ and\ \bibinfo {author} {\bibfnamefont {K.}~\bibnamefont {Gofryk}},\ }\bibfield  {title} {\bibinfo {title} {{Piezomagnetism and magnetoelastic memory in uranium dioxide}},\ }\href {https://doi.org/10.1038/s41467-017-00172-5} {\bibfield  {journal} {\bibinfo  {journal} {Nat. Commun}\ }\textbf {\bibinfo {volume} {8}},\ \bibinfo {pages} {99} (\bibinfo
  {year} {2017})}\BibitemShut {NoStop}%
\bibitem [{\citenamefont {Nanjo}\ \emph {et~al.}(2025)\citenamefont {Nanjo}, \citenamefont {Imai}, \citenamefont {Aoyama}, \citenamefont {Yamaura},\ and\ \citenamefont {Ohgushi}}]{nanjo2025piezomagneticeffect5dtransition}%
  \BibitemOpen
  \bibfield  {author} {\bibinfo {author} {\bibfnamefont {H.}~\bibnamefont {Nanjo}}, \bibinfo {author} {\bibfnamefont {Y.}~\bibnamefont {Imai}}, \bibinfo {author} {\bibfnamefont {T.}~\bibnamefont {Aoyama}}, \bibinfo {author} {\bibfnamefont {J.}~\bibnamefont {Yamaura}},\ and\ \bibinfo {author} {\bibfnamefont {K.}~\bibnamefont {Ohgushi}},\ }\href {https://arxiv.org/abs/2505.11145} {\bibinfo {title} {{Piezomagnetic effect in 5$d$ transition metal oxides Y$_2$Ir$_2$O$_7$ and Cd$_2$Os$_2$O$_7$ with all-in/all-out magnetic order}}} (\bibinfo {year} {2025}),\ \Eprint {https://arxiv.org/abs/2505.11145} {arXiv:2505.11145 [cond-mat.str-el]} \BibitemShut {NoStop}%
\bibitem [{\citenamefont {Tomikawa}\ \emph {et~al.}(2024)\citenamefont {Tomikawa}, \citenamefont {Araki}, \citenamefont {Ikeda}, \citenamefont {Nakamura}, \citenamefont {Aoki}, \citenamefont {Ishida},\ and\ \citenamefont {Yonezawa}}]{Tomikawa_PhysRevB.110.L100408}%
  \BibitemOpen
  \bibfield  {author} {\bibinfo {author} {\bibfnamefont {M.}~\bibnamefont {Tomikawa}}, \bibinfo {author} {\bibfnamefont {R.}~\bibnamefont {Araki}}, \bibinfo {author} {\bibfnamefont {A.}~\bibnamefont {Ikeda}}, \bibinfo {author} {\bibfnamefont {A.}~\bibnamefont {Nakamura}}, \bibinfo {author} {\bibfnamefont {D.}~\bibnamefont {Aoki}}, \bibinfo {author} {\bibfnamefont {K.}~\bibnamefont {Ishida}},\ and\ \bibinfo {author} {\bibfnamefont {S.}~\bibnamefont {Yonezawa}},\ }\bibfield  {title} {\bibinfo {title} {{Piezomagnetism in the Ising ferromagnet URhGe}},\ }\href {https://doi.org/10.1103/PhysRevB.110.L100408} {\bibfield  {journal} {\bibinfo  {journal} {Phys. Rev. B}\ }\textbf {\bibinfo {volume} {110}},\ \bibinfo {pages} {L100408} (\bibinfo {year} {2024})}\BibitemShut {NoStop}%
\bibitem [{\citenamefont {Ma}\ \emph {et~al.}(2021)\citenamefont {Ma}, \citenamefont {Hu}, \citenamefont {Li}, \citenamefont {Liu}, \citenamefont {Yao}, \citenamefont {Jia},\ and\ \citenamefont {Liu}}]{Ma_NatCommun2021}%
  \BibitemOpen
  \bibfield  {author} {\bibinfo {author} {\bibfnamefont {H.-Y.}\ \bibnamefont {Ma}}, \bibinfo {author} {\bibfnamefont {M.}~\bibnamefont {Hu}}, \bibinfo {author} {\bibfnamefont {N.}~\bibnamefont {Li}}, \bibinfo {author} {\bibfnamefont {J.}~\bibnamefont {Liu}}, \bibinfo {author} {\bibfnamefont {W.}~\bibnamefont {Yao}}, \bibinfo {author} {\bibfnamefont {J.-F.}\ \bibnamefont {Jia}},\ and\ \bibinfo {author} {\bibfnamefont {J.}~\bibnamefont {Liu}},\ }\bibfield  {title} {\bibinfo {title} {{Multifunctional antiferromagnetic materials with giant piezomagnetism and noncollinear spin current}},\ }\href {https://doi.org/10.1038/s41467-021-23127-7} {\bibfield  {journal} {\bibinfo  {journal} {Nat. Commun.}\ }\textbf {\bibinfo {volume} {12}},\ \bibinfo {pages} {2846} (\bibinfo {year} {2021})}\BibitemShut {NoStop}%
\bibitem [{\citenamefont {Bhowal}\ and\ \citenamefont {Spaldin}(2024)}]{Bhowal_PhysRevX.14.011019}%
  \BibitemOpen
  \bibfield  {author} {\bibinfo {author} {\bibfnamefont {S.}~\bibnamefont {Bhowal}}\ and\ \bibinfo {author} {\bibfnamefont {N.~A.}\ \bibnamefont {Spaldin}},\ }\bibfield  {title} {\bibinfo {title} {{Ferroically Ordered Magnetic Octupoles in $d$-Wave Altermagnets}},\ }\href {https://doi.org/10.1103/PhysRevX.14.011019} {\bibfield  {journal} {\bibinfo  {journal} {Phys. Rev. X}\ }\textbf {\bibinfo {volume} {14}},\ \bibinfo {pages} {011019} (\bibinfo {year} {2024})}\BibitemShut {NoStop}%
\bibitem [{\citenamefont {Naka}\ \emph {et~al.}(2025)\citenamefont {Naka}, \citenamefont {Motome}, \citenamefont {Miyazaki},\ and\ \citenamefont {Seo}}]{Naka2025}%
  \BibitemOpen
  \bibfield  {author} {\bibinfo {author} {\bibfnamefont {M.}~\bibnamefont {Naka}}, \bibinfo {author} {\bibfnamefont {Y.}~\bibnamefont {Motome}}, \bibinfo {author} {\bibfnamefont {T.}~\bibnamefont {Miyazaki}},\ and\ \bibinfo {author} {\bibfnamefont {H.}~\bibnamefont {Seo}},\ }\bibfield  {title} {\bibinfo {title} {{Nonrelativistic Piezomagnetic Effect in an Organic Altermagnet}},\ }\href {https://doi.org/10.7566/JPSJ.94.083702} {\bibfield  {journal} {\bibinfo  {journal} {J. Phys. Soc. Jpn.}\ }\textbf {\bibinfo {volume} {94}},\ \bibinfo {pages} {083702} (\bibinfo {year} {2025})}\BibitemShut {NoStop}%
\bibitem [{\citenamefont {Ogawa}\ and\ \citenamefont {Hayami}(2025)}]{Ogawa_2025}%
  \BibitemOpen
  \bibfield  {author} {\bibinfo {author} {\bibfnamefont {Y.}~\bibnamefont {Ogawa}}\ and\ \bibinfo {author} {\bibfnamefont {S.}~\bibnamefont {Hayami}},\ }\bibfield  {title} {\bibinfo {title} {{Nonlinear Piezomagnetic Effects in g-wave Altermagnets}},\ }\href {https://doi.org/10.7566/JPSJ.94.063704} {\bibfield  {journal} {\bibinfo  {journal} {J. Phys. Soc. Jpn.}\ }\textbf {\bibinfo {volume} {94}},\ \bibinfo {pages} {063704} (\bibinfo {year} {2025})}\BibitemShut {NoStop}%
\bibitem [{\citenamefont {Gomonaj}(1989)}]{Gomonaj_phasetransition1989}%
  \BibitemOpen
  \bibfield  {author} {\bibinfo {author} {\bibfnamefont {E.~V.}\ \bibnamefont {Gomonaj}},\ }\bibfield  {title} {\bibinfo {title} {{Magnetostriction and piezomagnetism of noncollinear antiferromagnet Mn3NiN}},\ }\href {https://doi.org/10.1080/01411598908206858} {\bibfield  {journal} {\bibinfo  {journal} {Phase Transitions}\ }\textbf {\bibinfo {volume} {18}},\ \bibinfo {pages} {93} (\bibinfo {year} {1989})}\BibitemShut {NoStop}%
\bibitem [{\citenamefont {Lukashev}\ \emph {et~al.}(2008)\citenamefont {Lukashev}, \citenamefont {Sabirianov},\ and\ \citenamefont {Belashchenko}}]{Lukashev2008}%
  \BibitemOpen
  \bibfield  {author} {\bibinfo {author} {\bibfnamefont {P.}~\bibnamefont {Lukashev}}, \bibinfo {author} {\bibfnamefont {R.~F.}\ \bibnamefont {Sabirianov}},\ and\ \bibinfo {author} {\bibfnamefont {K.}~\bibnamefont {Belashchenko}},\ }\bibfield  {title} {\bibinfo {title} {{Theory of the piezomagnetic effect in Mn-based antiperovskites}},\ }\href {https://doi.org/10.1103/PhysRevB.78.184414} {\bibfield  {journal} {\bibinfo  {journal} {Phys. Rev. B}\ }\textbf {\bibinfo {volume} {78}},\ \bibinfo {pages} {184414} (\bibinfo {year} {2008})}\BibitemShut {NoStop}%
\bibitem [{\citenamefont {Zemen}\ \emph {et~al.}(2017)\citenamefont {Zemen}, \citenamefont {Gercsi},\ and\ \citenamefont {Sandeman}}]{Zemen2017}%
  \BibitemOpen
  \bibfield  {author} {\bibinfo {author} {\bibfnamefont {J.}~\bibnamefont {Zemen}}, \bibinfo {author} {\bibfnamefont {Z.}~\bibnamefont {Gercsi}},\ and\ \bibinfo {author} {\bibfnamefont {K.}~\bibnamefont {Sandeman}},\ }\bibfield  {title} {\bibinfo {title} {{Piezomagnetism as a counterpart of the magnetovolume effect in magnetically frustrated Mn-based antiperovskite nitrides}},\ }\href {https://doi.org/10.1103/PhysRevB.96.024451} {\bibfield  {journal} {\bibinfo  {journal} {Phys. Rev. B}\ }\textbf {\bibinfo {volume} {96}},\ \bibinfo {pages} {024451} (\bibinfo {year} {2017})}\BibitemShut {NoStop}%
\bibitem [{\citenamefont {Boldrin}\ \emph {et~al.}(2018)\citenamefont {Boldrin}, \citenamefont {Mihai}, \citenamefont {Zou}, \citenamefont {Zemen}, \citenamefont {Thompson}, \citenamefont {Ware}, \citenamefont {Neamtu}, \citenamefont {Ghivelder}, \citenamefont {Esser}, \citenamefont {McComb}, \citenamefont {Petrov},\ and\ \citenamefont {Cohen}}]{Boldrin2018}%
  \BibitemOpen
  \bibfield  {author} {\bibinfo {author} {\bibfnamefont {D.}~\bibnamefont {Boldrin}}, \bibinfo {author} {\bibfnamefont {A.~P.}\ \bibnamefont {Mihai}}, \bibinfo {author} {\bibfnamefont {B.}~\bibnamefont {Zou}}, \bibinfo {author} {\bibfnamefont {J.}~\bibnamefont {Zemen}}, \bibinfo {author} {\bibfnamefont {R.}~\bibnamefont {Thompson}}, \bibinfo {author} {\bibfnamefont {E.}~\bibnamefont {Ware}}, \bibinfo {author} {\bibfnamefont {B.~V.}\ \bibnamefont {Neamtu}}, \bibinfo {author} {\bibfnamefont {L.}~\bibnamefont {Ghivelder}}, \bibinfo {author} {\bibfnamefont {B.}~\bibnamefont {Esser}}, \bibinfo {author} {\bibfnamefont {D.~W.}\ \bibnamefont {McComb}}, \bibinfo {author} {\bibfnamefont {P.}~\bibnamefont {Petrov}},\ and\ \bibinfo {author} {\bibfnamefont {L.~F.}\ \bibnamefont {Cohen}},\ }\bibfield  {title} {\bibinfo {title} {{Giant Piezomagnetism in Mn$_3$NiN}},\ }\href {https://doi.org/10.1021/acsami.8b03112} {\bibfield  {journal} {\bibinfo  {journal} {ACS Applied Materials \& Interfaces}\ }\textbf {\bibinfo {volume}
  {10}},\ \bibinfo {pages} {18863} (\bibinfo {year} {2018})}\BibitemShut {NoStop}%
\bibitem [{\citenamefont {Samathrakis}\ and\ \citenamefont {Zhang}(2020)}]{Samathrakis2020}%
  \BibitemOpen
  \bibfield  {author} {\bibinfo {author} {\bibfnamefont {I.}~\bibnamefont {Samathrakis}}\ and\ \bibinfo {author} {\bibfnamefont {H.}~\bibnamefont {Zhang}},\ }\bibfield  {title} {\bibinfo {title} {{Tailoring the anomalous Hall effect in the noncollinear antiperovskite $\mathrm{{Mn}}_{3}\mathrm{{G}a{N}}$}},\ }\href {https://doi.org/10.1103/PhysRevB.101.214423} {\bibfield  {journal} {\bibinfo  {journal} {Phys. Rev. B}\ }\textbf {\bibinfo {volume} {101}},\ \bibinfo {pages} {214423} (\bibinfo {year} {2020})}\BibitemShut {NoStop}%
\bibitem [{\citenamefont {Arima}(2013)}]{Arima_JPSJ2013}%
  \BibitemOpen
  \bibfield  {author} {\bibinfo {author} {\bibfnamefont {T.-h.}\ \bibnamefont {Arima}},\ }\bibfield  {title} {\bibinfo {title} {{Time-Reversal Symmetry Breaking and Consequent Physical Responses Induced by All-In-All-Out Type Magnetic Order on the Pyrochlore Lattice}},\ }\href {https://doi.org/10.7566/JPSJ.82.013705} {\bibfield  {journal} {\bibinfo  {journal} {J. Phys. Soc. Jpn.}\ }\textbf {\bibinfo {volume} {82}},\ \bibinfo {pages} {013705} (\bibinfo {year} {2013})}\BibitemShut {NoStop}%
\bibitem [{\citenamefont {Choi}\ \emph {et~al.}(2025)\citenamefont {Choi}, \citenamefont {Qualter},\ and\ \citenamefont {Bikouvaraki}}]{Choi2025}%
  \BibitemOpen
  \bibfield  {author} {\bibinfo {author} {\bibfnamefont {B.~C.}\ \bibnamefont {Choi}}, \bibinfo {author} {\bibfnamefont {J.}~\bibnamefont {Qualter}},\ and\ \bibinfo {author} {\bibfnamefont {K.}~\bibnamefont {Bikouvaraki}},\ }\bibfield  {title} {\bibinfo {title} {{Piezomagnetic effect on spin-orbit torque driven magnetization processes in ${\mathrm{Mn}}_{3}\mathrm{Sn}$ under epitaxial in-plane tensile strain: An atomistic spin modeling study}},\ }\href {https://doi.org/10.1103/pfpv-8d96} {\bibfield  {journal} {\bibinfo  {journal} {Phys. Rev. B}\ }\textbf {\bibinfo {volume} {112}},\ \bibinfo {pages} {094406} (\bibinfo {year} {2025})}\BibitemShut {NoStop}%
\bibitem [{\citenamefont {Huyen}\ \emph {et~al.}(2025)\citenamefont {Huyen}, \citenamefont {Yanagi},\ and\ \citenamefont {Suzuki}}]{Huyen2025}%
  \BibitemOpen
  \bibfield  {author} {\bibinfo {author} {\bibfnamefont {V.~T.~N.}\ \bibnamefont {Huyen}}, \bibinfo {author} {\bibfnamefont {Y.}~\bibnamefont {Yanagi}},\ and\ \bibinfo {author} {\bibfnamefont {M.-T.}\ \bibnamefont {Suzuki}},\ }\bibfield  {title} {\bibinfo {title} {{Anisotropic piezomagnetism in noncollinear antiferromagnets}},\ }\href {https://doi.org/10.1103/wcwq-bfhb} {\bibfield  {journal} {\bibinfo  {journal} {Phys. Rev. B}\ }\textbf {\bibinfo {volume} {112}},\ \bibinfo {pages} {104421} (\bibinfo {year} {2025})}\BibitemShut {NoStop}%
\bibitem [{\citenamefont {Meng}\ \emph {et~al.}(2024)\citenamefont {Meng}, \citenamefont {Dong}, \citenamefont {Nie}, \citenamefont {Xu}, \citenamefont {Wang}, \citenamefont {Jiang}, \citenamefont {Zuo}, \citenamefont {Zhang}, \citenamefont {Li}, \citenamefont {Zhu}, \citenamefont {Balents},\ and\ \citenamefont {Behnia}}]{Meng2025}%
  \BibitemOpen
  \bibfield  {author} {\bibinfo {author} {\bibfnamefont {Q.}~\bibnamefont {Meng}}, \bibinfo {author} {\bibfnamefont {J.}~\bibnamefont {Dong}}, \bibinfo {author} {\bibfnamefont {P.}~\bibnamefont {Nie}}, \bibinfo {author} {\bibfnamefont {L.}~\bibnamefont {Xu}}, \bibinfo {author} {\bibfnamefont {J.}~\bibnamefont {Wang}}, \bibinfo {author} {\bibfnamefont {S.}~\bibnamefont {Jiang}}, \bibinfo {author} {\bibfnamefont {H.}~\bibnamefont {Zuo}}, \bibinfo {author} {\bibfnamefont {J.}~\bibnamefont {Zhang}}, \bibinfo {author} {\bibfnamefont {X.}~\bibnamefont {Li}}, \bibinfo {author} {\bibfnamefont {Z.}~\bibnamefont {Zhu}}, \bibinfo {author} {\bibfnamefont {L.}~\bibnamefont {Balents}},\ and\ \bibinfo {author} {\bibfnamefont {K.}~\bibnamefont {Behnia}},\ }\bibfield  {title} {\bibinfo {title} {{Magnetostriction, piezomagnetism and domain nucleation in a Kagome antiferromagnet}},\ }\href {https://doi.org/10.1038/s41467-024-51268-y} {\bibfield  {journal} {\bibinfo  {journal} {Nat. Commun.}\ }\textbf {\bibinfo {volume} {15}},\
  \bibinfo {pages} {6921} (\bibinfo {year} {2024})}\BibitemShut {NoStop}%
\bibitem [{\citenamefont {Kresse}\ and\ \citenamefont {Hafner}(1993)}]{Kresse1993}%
  \BibitemOpen
  \bibfield  {author} {\bibinfo {author} {\bibfnamefont {G.}~\bibnamefont {Kresse}}\ and\ \bibinfo {author} {\bibfnamefont {J.}~\bibnamefont {Hafner}},\ }\bibfield  {title} {\bibinfo {title} {{Ab initio molecular dynamics for liquid metals}},\ }\href {https://doi.org/10.1103/PhysRevB.47.558} {\bibfield  {journal} {\bibinfo  {journal} {Phys. Rev. B}\ }\textbf {\bibinfo {volume} {47}},\ \bibinfo {pages} {558} (\bibinfo {year} {1993})}\BibitemShut {NoStop}%
\bibitem [{\citenamefont {Kresse}\ and\ \citenamefont {Hafner}(1994)}]{Kresse1994}%
  \BibitemOpen
  \bibfield  {author} {\bibinfo {author} {\bibfnamefont {G.}~\bibnamefont {Kresse}}\ and\ \bibinfo {author} {\bibfnamefont {J.}~\bibnamefont {Hafner}},\ }\bibfield  {title} {\bibinfo {title} {{Ab initio molecular dynamics for open-shell transition metals}},\ }\href {https://doi.org/10.1103/PhysRevB.49.14251} {\bibfield  {journal} {\bibinfo  {journal} {Phys. Rev. B}\ }\textbf {\bibinfo {volume} {49}},\ \bibinfo {pages} {14251} (\bibinfo {year} {1994})}\BibitemShut {NoStop}%
\bibitem [{\citenamefont {Kresse}\ and\ \citenamefont {Furthm^^c3^^bcller}(1996{\natexlab{a}})}]{Kresse1996}%
  \BibitemOpen
  \bibfield  {author} {\bibinfo {author} {\bibfnamefont {G.}~\bibnamefont {Kresse}}\ and\ \bibinfo {author} {\bibfnamefont {J.}~\bibnamefont {Furthm^^c3^^bcller}},\ }\bibfield  {title} {\bibinfo {title} {{Efficient iterative schemes for ab initio total-energy calculations using a plane-wave basis set}},\ }\href {https://doi.org/10.1103/PhysRevB.54.11169} {\bibfield  {journal} {\bibinfo  {journal} {Phys. Rev. B}\ }\textbf {\bibinfo {volume} {54}},\ \bibinfo {pages} {11169} (\bibinfo {year} {1996}{\natexlab{a}})}\BibitemShut {NoStop}%
\bibitem [{\citenamefont {Kresse}\ and\ \citenamefont {Furthm^^c3^^bcller}(1996{\natexlab{b}})}]{Kresse1996CMS}%
  \BibitemOpen
  \bibfield  {author} {\bibinfo {author} {\bibfnamefont {G.}~\bibnamefont {Kresse}}\ and\ \bibinfo {author} {\bibfnamefont {J.}~\bibnamefont {Furthm^^c3^^bcller}},\ }\bibfield  {title} {\bibinfo {title} {{Efficiency of ab-initio total energy calculations for metals and semiconductors using a plane-wave basis set}},\ }\href {https://doi.org/10.1016/0927-0256(96)00008-0} {\bibfield  {journal} {\bibinfo  {journal} {Comput. Mater. Sci.}\ }\textbf {\bibinfo {volume} {6}},\ \bibinfo {pages} {15} (\bibinfo {year} {1996}{\natexlab{b}})}\BibitemShut {NoStop}%
\bibitem [{\citenamefont {Perdew}\ \emph {et~al.}(1996)\citenamefont {Perdew}, \citenamefont {Burke},\ and\ \citenamefont {Ernzerhof}}]{Perdew1996}%
  \BibitemOpen
  \bibfield  {author} {\bibinfo {author} {\bibfnamefont {J.~P.}\ \bibnamefont {Perdew}}, \bibinfo {author} {\bibfnamefont {K.}~\bibnamefont {Burke}},\ and\ \bibinfo {author} {\bibfnamefont {M.}~\bibnamefont {Ernzerhof}},\ }\bibfield  {title} {\bibinfo {title} {{Generalized Gradient Approximation Made Simple}},\ }\href {https://doi.org/10.1103/PhysRevLett.77.3865} {\bibfield  {journal} {\bibinfo  {journal} {Phys. Rev. Lett.}\ }\textbf {\bibinfo {volume} {77}},\ \bibinfo {pages} {3865} (\bibinfo {year} {1996})}\BibitemShut {NoStop}%
\bibitem [{\citenamefont {Bradley}\ and\ \citenamefont {Cracknell}(2010)}]{Bradley_book_2010}%
  \BibitemOpen
  \bibfield  {author} {\bibinfo {author} {\bibfnamefont {C.}~\bibnamefont {Bradley}}\ and\ \bibinfo {author} {\bibfnamefont {A.}~\bibnamefont {Cracknell}},\ }\href@noop {} {\emph {\bibinfo {title} {{The mathematical theory of symmetry in solids: representation theory for point groups and space groups}}}}\ (\bibinfo  {publisher} {Oxford University Press},\ \bibinfo {year} {2010})\BibitemShut {NoStop}%
\bibitem [{\citenamefont {Kawamura}(2019)}]{Kawamurafermisurf}%
  \BibitemOpen
  \bibfield  {author} {\bibinfo {author} {\bibfnamefont {M.}~\bibnamefont {Kawamura}},\ }\bibfield  {title} {\bibinfo {title} {{FermiSurfer: Fermi-surface viewer providing multiple representation schemes}},\ }\href {https://doi.org/10.1016/j.cpc.2019.01.017} {\bibfield  {journal} {\bibinfo  {journal} {Comp. Phys. Commun.}\ }\textbf {\bibinfo {volume} {239}},\ \bibinfo {pages} {197} (\bibinfo {year} {2019})}\BibitemShut {NoStop}%
\bibitem [{\citenamefont {Brinkman}\ and\ \citenamefont {Elliott}(1966)}]{Brinkman_SSG_1966}%
  \BibitemOpen
  \bibfield  {author} {\bibinfo {author} {\bibfnamefont {W.~F.}\ \bibnamefont {Brinkman}}\ and\ \bibinfo {author} {\bibfnamefont {R.~J.}\ \bibnamefont {Elliott}},\ }\bibfield  {title} {\bibinfo {title} {{Theory of Spin-Space Groups}},\ }\href@noop {} {\bibfield  {journal} {\bibinfo  {journal} {Proceedings of the Royal Society of London. Series A, Mathematical and Physical Sciences}\ }\textbf {\bibinfo {volume} {294}},\ \bibinfo {pages} {343} (\bibinfo {year} {1966})}\BibitemShut {NoStop}%
\bibitem [{\citenamefont {Litvin}\ and\ \citenamefont {Opechowski}(1974)}]{LITVIN1974538}%
  \BibitemOpen
  \bibfield  {author} {\bibinfo {author} {\bibfnamefont {D.~B.}\ \bibnamefont {Litvin}}\ and\ \bibinfo {author} {\bibfnamefont {W.}~\bibnamefont {Opechowski}},\ }\bibfield  {title} {\bibinfo {title} {{Spin groups}},\ }\href {https://doi.org/https://doi.org/10.1016/0031-8914(74)90157-8} {\bibfield  {journal} {\bibinfo  {journal} {Physica}\ }\textbf {\bibinfo {volume} {76}},\ \bibinfo {pages} {538} (\bibinfo {year} {1974})}\BibitemShut {NoStop}%
\bibitem [{\citenamefont {Litvin}(1977)}]{Litvin:a14103}%
  \BibitemOpen
  \bibfield  {author} {\bibinfo {author} {\bibfnamefont {D.~B.}\ \bibnamefont {Litvin}},\ }\bibfield  {title} {\bibinfo {title} {{{Spin point groups}}},\ }\href {https://doi.org/10.1107/S0567739477000709} {\bibfield  {journal} {\bibinfo  {journal} {Acta Crystallographica Section A}\ }\textbf {\bibinfo {volume} {33}},\ \bibinfo {pages} {279} (\bibinfo {year} {1977})}\BibitemShut {NoStop}%
\bibitem [{\citenamefont {Liu}\ \emph {et~al.}(2022)\citenamefont {Liu}, \citenamefont {Li}, \citenamefont {Han}, \citenamefont {Wan},\ and\ \citenamefont {Liu}}]{Liu_PhysRevX.12.021016}%
  \BibitemOpen
  \bibfield  {author} {\bibinfo {author} {\bibfnamefont {P.}~\bibnamefont {Liu}}, \bibinfo {author} {\bibfnamefont {J.}~\bibnamefont {Li}}, \bibinfo {author} {\bibfnamefont {J.}~\bibnamefont {Han}}, \bibinfo {author} {\bibfnamefont {X.}~\bibnamefont {Wan}},\ and\ \bibinfo {author} {\bibfnamefont {Q.}~\bibnamefont {Liu}},\ }\bibfield  {title} {\bibinfo {title} {{Spin-Group Symmetry in Magnetic Materials with Negligible Spin-Orbit Coupling}},\ }\href {https://doi.org/10.1103/PhysRevX.12.021016} {\bibfield  {journal} {\bibinfo  {journal} {Phys. Rev. X}\ }\textbf {\bibinfo {volume} {12}},\ \bibinfo {pages} {021016} (\bibinfo {year} {2022})}\BibitemShut {NoStop}%
\bibitem [{\citenamefont {\ifmmode~\check{S}\else \v{S}\fi{}mejkal}\ \emph {et~al.}(2022)\citenamefont {\ifmmode~\check{S}\else \v{S}\fi{}mejkal}, \citenamefont {Sinova},\ and\ \citenamefont {Jungwirth}}]{Smejkal_PhysRevX.12.031042}%
  \BibitemOpen
  \bibfield  {author} {\bibinfo {author} {\bibfnamefont {L.}~\bibnamefont {\ifmmode~\check{S}\else \v{S}\fi{}mejkal}}, \bibinfo {author} {\bibfnamefont {J.}~\bibnamefont {Sinova}},\ and\ \bibinfo {author} {\bibfnamefont {T.}~\bibnamefont {Jungwirth}},\ }\bibfield  {title} {\bibinfo {title} {{Beyond Conventional Ferromagnetism and Antiferromagnetism: A Phase with Nonrelativistic Spin and Crystal Rotation Symmetry}},\ }\href {https://doi.org/10.1103/PhysRevX.12.031042} {\bibfield  {journal} {\bibinfo  {journal} {Phys. Rev. X}\ }\textbf {\bibinfo {volume} {12}},\ \bibinfo {pages} {031042} (\bibinfo {year} {2022})}\BibitemShut {NoStop}%
\bibitem [{\citenamefont {Watanabe}\ \emph {et~al.}(2024)\citenamefont {Watanabe}, \citenamefont {Shinohara}, \citenamefont {Nomoto}, \citenamefont {Togo},\ and\ \citenamefont {Arita}}]{Watanabe_PhysRevB.109.094438}%
  \BibitemOpen
  \bibfield  {author} {\bibinfo {author} {\bibfnamefont {H.}~\bibnamefont {Watanabe}}, \bibinfo {author} {\bibfnamefont {K.}~\bibnamefont {Shinohara}}, \bibinfo {author} {\bibfnamefont {T.}~\bibnamefont {Nomoto}}, \bibinfo {author} {\bibfnamefont {A.}~\bibnamefont {Togo}},\ and\ \bibinfo {author} {\bibfnamefont {R.}~\bibnamefont {Arita}},\ }\bibfield  {title} {\bibinfo {title} {{Symmetry analysis with spin crystallographic groups: Disentangling effects free of spin-orbit coupling in emergent electromagnetism}},\ }\href {https://doi.org/10.1103/PhysRevB.109.094438} {\bibfield  {journal} {\bibinfo  {journal} {Phys. Rev. B}\ }\textbf {\bibinfo {volume} {109}},\ \bibinfo {pages} {094438} (\bibinfo {year} {2024})}\BibitemShut {NoStop}%
\bibitem [{\citenamefont {Xiao}\ \emph {et~al.}(2024)\citenamefont {Xiao}, \citenamefont {Zhao}, \citenamefont {Li}, \citenamefont {Shindou},\ and\ \citenamefont {Song}}]{Zhenyu_PhysRevX.14.031037}%
  \BibitemOpen
  \bibfield  {author} {\bibinfo {author} {\bibfnamefont {Z.}~\bibnamefont {Xiao}}, \bibinfo {author} {\bibfnamefont {J.}~\bibnamefont {Zhao}}, \bibinfo {author} {\bibfnamefont {Y.}~\bibnamefont {Li}}, \bibinfo {author} {\bibfnamefont {R.}~\bibnamefont {Shindou}},\ and\ \bibinfo {author} {\bibfnamefont {Z.-D.}\ \bibnamefont {Song}},\ }\bibfield  {title} {\bibinfo {title} {{Spin Space Groups: Full Classification and Applications}},\ }\href {https://doi.org/10.1103/PhysRevX.14.031037} {\bibfield  {journal} {\bibinfo  {journal} {Phys. Rev. X}\ }\textbf {\bibinfo {volume} {14}},\ \bibinfo {pages} {031037} (\bibinfo {year} {2024})}\BibitemShut {NoStop}%
\bibitem [{\citenamefont {Chen}\ \emph {et~al.}(2024)\citenamefont {Chen}, \citenamefont {Ren}, \citenamefont {Zhu}, \citenamefont {Yu}, \citenamefont {Zhang}, \citenamefont {Liu}, \citenamefont {Li}, \citenamefont {Liu}, \citenamefont {Li},\ and\ \citenamefont {Liu}}]{Xiaobing_PhysRevX.14.031038}%
  \BibitemOpen
  \bibfield  {author} {\bibinfo {author} {\bibfnamefont {X.}~\bibnamefont {Chen}}, \bibinfo {author} {\bibfnamefont {J.}~\bibnamefont {Ren}}, \bibinfo {author} {\bibfnamefont {Y.}~\bibnamefont {Zhu}}, \bibinfo {author} {\bibfnamefont {Y.}~\bibnamefont {Yu}}, \bibinfo {author} {\bibfnamefont {A.}~\bibnamefont {Zhang}}, \bibinfo {author} {\bibfnamefont {P.}~\bibnamefont {Liu}}, \bibinfo {author} {\bibfnamefont {J.}~\bibnamefont {Li}}, \bibinfo {author} {\bibfnamefont {Y.}~\bibnamefont {Liu}}, \bibinfo {author} {\bibfnamefont {C.}~\bibnamefont {Li}},\ and\ \bibinfo {author} {\bibfnamefont {Q.}~\bibnamefont {Liu}},\ }\bibfield  {title} {\bibinfo {title} {{Enumeration and Representation Theory of Spin Space Groups}},\ }\href {https://doi.org/10.1103/PhysRevX.14.031038} {\bibfield  {journal} {\bibinfo  {journal} {Phys. Rev. X}\ }\textbf {\bibinfo {volume} {14}},\ \bibinfo {pages} {031038} (\bibinfo {year} {2024})}\BibitemShut {NoStop}%
\bibitem [{\citenamefont {Jiang}\ \emph {et~al.}(2024)\citenamefont {Jiang}, \citenamefont {Song}, \citenamefont {Zhu}, \citenamefont {Fang}, \citenamefont {Weng}, \citenamefont {Liu}, \citenamefont {Yang},\ and\ \citenamefont {Fang}}]{Jiang_PhysRevX.14.031039}%
  \BibitemOpen
  \bibfield  {author} {\bibinfo {author} {\bibfnamefont {Y.}~\bibnamefont {Jiang}}, \bibinfo {author} {\bibfnamefont {Z.}~\bibnamefont {Song}}, \bibinfo {author} {\bibfnamefont {T.}~\bibnamefont {Zhu}}, \bibinfo {author} {\bibfnamefont {Z.}~\bibnamefont {Fang}}, \bibinfo {author} {\bibfnamefont {H.}~\bibnamefont {Weng}}, \bibinfo {author} {\bibfnamefont {Z.-X.}\ \bibnamefont {Liu}}, \bibinfo {author} {\bibfnamefont {J.}~\bibnamefont {Yang}},\ and\ \bibinfo {author} {\bibfnamefont {C.}~\bibnamefont {Fang}},\ }\bibfield  {title} {\bibinfo {title} {{Enumeration of Spin-Space Groups: Toward a Complete Description of Symmetries of Magnetic Orders}},\ }\href {https://doi.org/10.1103/PhysRevX.14.031039} {\bibfield  {journal} {\bibinfo  {journal} {Phys. Rev. X}\ }\textbf {\bibinfo {volume} {14}},\ \bibinfo {pages} {031039} (\bibinfo {year} {2024})}\BibitemShut {NoStop}%
\bibitem [{\citenamefont {Schiff}\ \emph {et~al.}(2025)\citenamefont {Schiff}, \citenamefont {Corticelli}, \citenamefont {Guerreiro}, \citenamefont {Romh^^c3^^a1nyi},\ and\ \citenamefont {McClarty}}]{Schiff_SciPostPhys.18.3.109}%
  \BibitemOpen
  \bibfield  {author} {\bibinfo {author} {\bibfnamefont {H.}~\bibnamefont {Schiff}}, \bibinfo {author} {\bibfnamefont {A.}~\bibnamefont {Corticelli}}, \bibinfo {author} {\bibfnamefont {A.}~\bibnamefont {Guerreiro}}, \bibinfo {author} {\bibfnamefont {J.}~\bibnamefont {Romh^^c3^^a1nyi}},\ and\ \bibinfo {author} {\bibfnamefont {P.}~\bibnamefont {McClarty}},\ }\bibfield  {title} {\bibinfo {title} {{The crystallographic spin point groups and their representations}},\ }\href {https://doi.org/10.21468/SciPostPhys.18.3.109} {\bibfield  {journal} {\bibinfo  {journal} {SciPost Phys.}\ }\textbf {\bibinfo {volume} {18}},\ \bibinfo {pages} {109} (\bibinfo {year} {2025})}\BibitemShut {NoStop}%
\bibitem [{\citenamefont {Song}\ \emph {et~al.}(2025)\citenamefont {Song}, \citenamefont {Yang}, \citenamefont {Jiang}, \citenamefont {Fang}, \citenamefont {Yang}, \citenamefont {Fang}, \citenamefont {Weng},\ and\ \citenamefont {Liu}}]{Song_PhysRevB.111.134407}%
  \BibitemOpen
  \bibfield  {author} {\bibinfo {author} {\bibfnamefont {Z.}~\bibnamefont {Song}}, \bibinfo {author} {\bibfnamefont {A.~Z.}\ \bibnamefont {Yang}}, \bibinfo {author} {\bibfnamefont {Y.}~\bibnamefont {Jiang}}, \bibinfo {author} {\bibfnamefont {Z.}~\bibnamefont {Fang}}, \bibinfo {author} {\bibfnamefont {J.}~\bibnamefont {Yang}}, \bibinfo {author} {\bibfnamefont {C.}~\bibnamefont {Fang}}, \bibinfo {author} {\bibfnamefont {H.}~\bibnamefont {Weng}},\ and\ \bibinfo {author} {\bibfnamefont {Z.-X.}\ \bibnamefont {Liu}},\ }\bibfield  {title} {\bibinfo {title} {{Constructions and applications of irreducible representations of spin-space groups}},\ }\href {https://doi.org/10.1103/PhysRevB.111.134407} {\bibfield  {journal} {\bibinfo  {journal} {Phys. Rev. B}\ }\textbf {\bibinfo {volume} {111}},\ \bibinfo {pages} {134407} (\bibinfo {year} {2025})}\BibitemShut {NoStop}%
\end{thebibliography}%

\end{document}